\documentclass[aps,prd,twocolumn,eqsecnum,amssymb,amsmath,showpacs,a4paper, superscriptaddress]{revtex4-2}

\usepackage{graphicx}
\usepackage{amsfonts}
\usepackage{amsmath}
\usepackage{hyperref}
\usepackage{units}
\usepackage{color}
\usepackage{dcolumn}
\usepackage{bm}
\usepackage{mathrsfs}
\usepackage{float}
\usepackage{cleveref}
\usepackage[normalem]{ulem}
\usepackage{appendix}
\usepackage{mathtools}
\usepackage{esvect}
\usepackage[normalem]{ulem}

\usepackage{array,multirow,graphicx}

\newcommand{\m}{\widetilde{m}}
\newcommand{\tp}{{tp}}
\newcommand{\cst}{\mathrm{const}}
\newcommand{\sgn}{\mathrm{sgn}}
\newcommand{\Afr}{\mathcal{A}}

\begin{document}

\title{Quantum vortex instability and black hole superradiance}
\author{Sam Patrick}
\affiliation{Department of Physics and Astronomy, University of British Columbia, Vancouver, British Columbia, V6T 1Z1, Canada}
\affiliation{Institute of Quantum Science and Engineering, Texas A\&M University, College Station, Texas, 77840, US}

\author{August Geelmuyden}
\affiliation{School of Mathematical Sciences, University of Nottingham, University Park, Nottingham, NG7 2RD, UK}

\author{Sebastian Erne}
\affiliation{Vienna Center for Quantum Science and Technology, Atominstitut, TU Wien, Stadionallee 2, 1020 Vienna, Austria}

\author{Carlo F. Barenghi}
\affiliation{Joint Quantum Centre Durham-Newcastle, School of Mathematics, Statistics and Physics, Newcastle University,
Newcastle upon Tyne, NE1 7RU, UK}

\author{Silke Weinfurtner}
\affiliation{School of Mathematical Sciences, University of Nottingham, University Park, Nottingham, NG7 2RD, UK}
\affiliation{Centre for the Mathematics and Theoretical Physics of Quantum Non-Equilibrium Systems, University of Nottingham, Nottingham, NG7 2RD, UK}

\date{\today}

\begin{abstract}
Vortices and black holes set the scene for many interesting dynamical processes in physics.
Here, we study the dynamical instability of quantised vortices and rotational superradiance around rotating black holes, illustrating in the process that the same physics is at play in these two seemingly disparate phenomena.
We also compare the instability of the vortex to the black hole bomb instability, which occurs for massive scalar fields in the Kerr spacetime.
Taking inspiration from the analogy between black hole bomb modes and the hydrogen spectrum, the vortex instability is compared with nuclear resonances involved in $\alpha$-decay.
\end{abstract}

\maketitle

\section{Introduction}

It is a remarkable fact that phenomena in completely separate areas of physics can admit the same theoretical description.
Historically, the pervasiveness of certain descriptions has served not only as a tool to aid our understanding of known phenomena, but also as a driving force in the development of new theories.
A prominent example is Maxwell's use of the fluid equations as a ``vehicle of mathematical reasoning'' in his development of classical electrodynamics \cite{maxwell1890scientific}.
Another example is Bekenstein's identification of the similarity between the laws of thermodynamics and black hole (BH) physics \cite{bekenstein1973black,bekenstein1974generalised}, which ultimately led to Hawking's discovery of black hole evaporation \cite{hawking1974explosions}.
Theoretical similarities are often indicative that the same conceptual mechanism underpins different physical processes.
In this paper, we investigate such a similarity between the instability of quantum vortices in superfluids and the superradiant amplification of waves around rotating BHs.

In recent years, there has been a growing body of work in the literature drawing comparisons between hydrodynamical vortices and rotating BHs \cite{basak2003superresonance,basak2003reflection,berti2004qnm,basak2005analog,slatyer2005superradiant,slatyer2005superradiant,federici2006superradiance,richartz2015rotating,churilov2018scattering,demirkaya2020acoustic}.
This is largely due to the success of the analogue gravity programme, which aims to shed light on processes that occur for (quantum) fields in curved spacetime by studying their counterparts in condensed matter physics \cite{barcelo2011analogue}.
The basic notion at the heart of analogue gravity is that the excitations of many fluid-like systems obey (in the non-dispersive regime) an equation which is formally equivalent to that of a Klein-Gordon field in a curved spacetime \cite{unruh1981experimental}.
This correspondence has led to the detection of analogue Hawking radiation 
\cite{weinfurtner2011measurement,euve2016observation,steinhauer2016observation,denova2019observation,kolobov2019spontaneous}
and superradiance \cite{torres2017rotational} in the laboratory.
Here, we use the term superradiance to refer to rotational superradiance (commonly studied in BH physics \cite{brito2020superradiance}) as distinct from Dicke superradiance in quantum optics \cite{dicke1954coherence}.
In view of the experiments of \cite{torres2017rotational}, which were performed in the dispersive regime, two of the authors have theoretically demonstrated that superradiance persists well-beyond the regime of the correspondence with the Klein-Gordon equation \cite{patrick2020superradiance,patrick2021rotational}, highlighting the possibility of studying BH processes in a much wider range of scenarios than previously thought possible.
The mechanism which enables superradiance is the existence of a region where excitations of the system can have negative energies. Such a region is commonly called an ergoregion, as it is a region where excitations do work on the system to extract energy from it.
The negative energy of such excitations then simply captures the fact that the energy of the system is being lowered.

A promising system within which to pursue this analogy involves flowing Bose-Einstein condensates (BECs).
One of the most striking features of BECs (and superfluids in general) is that circulation is quantised \cite{fetter2001vortices}.
This has been known since the predictions of Onsager and Feynman in the mid-20th century \cite{feynman1955application,onsager1949statistical} and has been researched heavily in the decades since, see e.g. \cite{fetter2009rotating} and references therein.
From the BH perspective, these systems are interesting since they provide a way to simulate effective spacetimes with quantum features.
From the condensed matter point of view, the analogy with BHs is useful since it can offer new perspectives on known BEC phenomena.

It has long been recognised that multiply-quantised (also described as highly-charged or large) vortices are unstable and will decay into clusters of singly-quantised vortices.
This decay occurs in dissipative systems since splitting into a cluster lowers the total energy \cite{barenghi2016primer}, however, it can also occur in non-dissipative systems if the vortex is able to transfer energy into sound waves \cite{takeuchi2018doubly}.
This instability has been confirmed both numerically and experimentally \cite{shin2004dynamical,isoshima2007spontaneous,okano2007splitting}.
Interestingly, it has been argued that multiply-quantised vortices in condensates can be stabilised by a draining component in the flow field \cite{zezyulin2014stationary,alperin2021multiply}.
In general, a deeper understanding of vortex dynamics is desirable due to the implications for	 on-going research on (e.g.) wave-vortex interactions \cite{geelmuyden2021sound} and quantum turbulence \cite{johnstone2019evolution}.

The connection between vortex instabilities and superradiance (or more precisely ergoregions) has been alluded to on several occasions in the literature.
Early studies focussed on the non-dispersive regime and ad hoc boundary conditions (BCs) were imposed near the vortex axis, where the fluid density drops to zero, to circumvent the breakdown of the hydrodynamic approximation there \cite{oliveira2014ergoregion,oliveira2018ergoregion}.
These works described this instability as an ergoregion instability, a term coined in the 1970s for describing instabilities of rapidly rotating neutron stars \cite{schutz1975gravitational}.
The ergoregion instability has also been referred to as an ``inverted BH bomb'' \cite{cardoso2008ergoregion}, where the BH bomb is a similar superradiant instability which results from the trapping of amplified modes outside the ergoregion. This occurs, for example, for massive scalar fields in the Kerr spacetime \cite{dolan2007instability}.
Recently, it has been shown in \cite{giacomelli2020ergoregion} that the link between the ergoregion and vortex instabilities also carries over to the dispersive regime.
Specifically, \cite{giacomelli2020ergoregion} demonstrated that the vortex instability originates in the vortex core where the mode has negative energy, thereby identifying the phenomenon as an ergoregion instability and not a BH bomb.
In this work, we provide further insight into the link between vortex instabilities, ergoregions and superradiance by illustrating how each can be analysed within the same framework (introduced below).
We also perform an explicit comparison between the ergoregion and BH bomb instabilities.

\section{Methods}

Our main tool to study the physics underlying vortex instabilities and superradiance will be based on the dispersion relation.
To obtain the dispersion relation, one assumes that the solutions to the linear equations of motion look locally like plane waves, which amounts to a Wenzel-Kramers-Brillouin (WKB) approximation.
The essence of the WKB approximation is that the waves vary rapidly when compared with the changes in the background they move through, allowing one (to leading order) to obtain a relation between the local wavelength and the frequency \cite{berry1972semiclassical,buhler2014waves}.
At next to leading order, one obtains a correction to this plane wave approximation in the form of a spatially varying amplitude.
At certain locations in the system, the amplitude diverges and the approximation breaks down, which signals the interaction of the different modes of the dispersion relation.
By solving the wave equation locally and matching asymptotically onto the WKB modes, one obtains a transfer matrix which relates the mode amplitudes in different regions of the system.
Armed with the transfer matrices, it is possible to estimate scattering coefficients and resonant frequencies.

The benefit of this method is that (1) one has precise control of the locations in the system where scattering and trapping of the waves take place, (2) it can be applied with relative ease even when conventional exact methods fail and (3) it converges to the exact solution in the limit of vanishingly small wavelengths (which corresponds in this work to high frequencies).
The first point is appealing to us here since the WKB method will allow us to pinpoint precisely where the trapping of unstable modes occurs.
The second is also important since BH instabilities are usually computed using a continued fraction algorithm \cite{dolan2007instability} which, as of yet, has not been reformulated to apply to dispersive systems.
The WKB method, however, can easily be adapted to handle dispersion, e.g. \cite{torres2018waves,patrick2020quasinormal,patrick2020superradiance,patrick2021rotational}.
The final point is relevant since the convergence of WKB to the exact solution often happens so quickly that approximation yields acceptable results even for relatively long wavelength modes.
Indeed, we will see examples in this work where WKB results agree with more accurate methods to within a couple of percent, even for low frequencies.

\section{Outline and main results} 

We now present an outline of the paper along with a summary of our main results.
In Section~\ref{sec:vortex}, we study the instability of multiply-quantised vortices.
After establishing the preliminaries, we show how the WKB method can be applied to study wave scattering in Section~\ref{sec:wkb_method}.
Then, in Section~\ref{sec:vortex_open}, we compute the unstable frequencies of a vortex in an infinite system using the WKB method.
This analysis reveals that the instability is a superradiant bound state in the ergoregion, i.e. a trapped negative energy excitation. Equivalently, one can think of this as a trapped positive frequency mode with negative norm (the norm is the particle number) which turns out to be a more useful notion for our discussion.
The validity of these results is confirmed by a comparison with full numerical simulations of the linearised equations.
Since real experiments are performed in finite sized systems, we show in Section~\ref{sec:vortex_close} how our WKB formula for the instabilities is modified when a reflective boundary is placed far from the vortex core.
An important result here is that the WKB instability condition reduces exactly to the open system condition when taking the outer boundary to infinity, thereby validating existing approaches in the literature \cite{takeuchi2018doubly,giacomelli2020ergoregion} which inferred stability properties of open systems by considering limits of closed systems.
In our companion paper \cite{patrick2021origin}, we study the instability in the non-linear regime and observe the subsequent decay of a doubly-quantised vortex into a pair of phase singularities. There, we also uncover novel behaviour of the system at late times: the system does not completely decay into two well-separated vortices but instead forms a bound state of phase defects orbiting in close proximity, whilst the separation between them undergoes a series of modulations.

Having identified the mechanism underpinning the vortex instabilities, next we study a non-dispersive draining vortex in Section~\ref{sec:dbt_vortex}, which is an analogue rotating BH.
This serves three main purposes.
Firstly, we demonstrate in Section~\ref{sec:dbt_wkb} that BH superradiance is also the result of positive frequency modes tunnelling to a region where they have negative norm, i.e. the same process that causes the vortex instabilities.
Secondly, the analogy to BHs allows us to identify the negative norm region of the system as the ergoregion, thereby justifying the name ergoregion instability for the unstable vortex modes.
Thirdly, we show in Section~\ref{sec:drain_stab} that the addition of a drain to a vortex in an infinite system can stabilise the system (at least in the non-dispersive regime).

Finally, we study the BH bomb instability of a massive scalar field in the Kerr spacetime.
After introducing the system in Section~\ref{sec:kerr}, in Section \ref{sec:kerr_BS} we apply the WKB method to demonstrate the existence of instabilities, showing in the process how our method can be used to recover the well-known hydrogenic approximation for the instability spectrum \cite{dolan2007instability}. 
We then explicitly compare the two confinement mechanisms for the BH bomb and vortex/ergoregion instabilities.
Taking inspiration from the hydrogen analogy, we conclude by arguing that the vortex instability shares common features with the nuclear resonances involved in $\alpha$-decay. 

\section{Quantised vortices} \label{sec:vortex}

\subsection{Gross-Pitaevskii equation} \label{sec:GPE}

Consider a two-dimensional condensate of bosons with nearly zero temperature. 
In the limit of many bosons, the system is well-described by the mean-field condensate wavefunction $\Psi(t,\mathbf{x})$, where $t$ is time and $\mathbf{x}$ are spatial coordinates on a 2D plane \cite{fetter2001vortices}.
The action for the system can be written as,
\begin{equation} \label{Sgpe1}
\begin{split}
\mathcal{S} = & \ \int dt d^2\mathbf{x} \bigg[ \frac{i\hbar}{2}\left(\Psi^*\partial_t\Psi-\Psi\partial_t\Psi^*\right) \\
& \ \ -\frac{\hbar^2}{2M}\bm{\nabla}\Psi\cdot\bm{\nabla}\Psi^* -U(\mathbf{x})|\Psi|^2 -\frac{1}{2}g|\Psi|^4\bigg],
\end{split}
\end{equation}
where $g$ is the 2D interaction parameter, $M$ is the mass of bosons in the condensate and $U$ is the trapping potential which depends on the spatial coordinate.
The square of the wavefunction has the interpretation of the particle density $\rho=|\Psi|^2$ and its integral gives the total number of atoms in the condensate, i.e. $N=\int d^2\mathbf{x} |\Psi|^2$.
In the ground state, the condensate oscillates at the chemical potential $\mu$, i.e. $\Psi\sim e^{-i\mu t}$, where $\mu$ sets the typical energy scale of the system.
To lighten notation, we can define the following parameters,
\begin{equation} \label{scales}
\xi = \hbar/\sqrt{M\mu}, \qquad \tau = \hbar/\mu, \qquad \rho_c = \mu/g,
\end{equation}
where $\xi$ is the healing length, $\tau$ is the healing time and $\rho_c$ is a reference density.
We then rescale all quantities in \eqref{Sgpe1} by these parameters.
Setting $\mathbf{x}/\xi\to\mathbf{x}$, $t/\tau\to t$, $\Psi/\sqrt{\rho_c}\to \Psi$, $U/\mu \to U$ and $\mathcal{S}/\hbar\to\mathcal{S}$, which amounts to setting $\hbar=M=\mu=g=1$, we obtain the action in dimensionless form,
\begin{equation} \label{Sgpe2}
\begin{split}
\mathcal{S} = & \ \int dt d^2\mathbf{x} \bigg[ \frac{i}{2}\left(\Psi^*\partial_t\Psi-\Psi\partial_t\Psi^*\right) \\ & \qquad -\frac{1}{2}\bm{\nabla}\Psi\cdot\bm{\nabla}\Psi^* -U(\mathbf{x})|\Psi|^2 -\frac{1}{2}|\Psi|^4\bigg].
\end{split}
\end{equation}
Variation of $\mathcal{S}$ with respect to $\Psi$ yields the dimensionless Gross-Pitaevskii equation (GPE),
\begin{equation} \label{GPE}
i\partial_t\Psi = -\frac{1}{2}\nabla^2\Psi + \left[U(\mathbf{x})+|\Psi|^2\right]\Psi,
\end{equation}
which is a non-linear Schr\"odinger equation. 
Instead varying \eqref{Sgpe2} with respect to $\Psi^*$, i.e. the complex conjugate of the wavefunction, yields the complex conjugate of \eqref{GPE}.

\subsection{Hydrodynamic variables} \label{sec:hydro}

The GPE can be put into hydrodynamic form by writing $\Psi$ in the Madelung representation,
\begin{equation}
\Psi = \sqrt{\rho}e^{i\Phi}.
\end{equation}
The GPE then splits into two separate equations,
\begin{align}
\partial_t\Phi + \frac{1}{2}(\bm{\nabla}\Phi)^2 + \rho - \frac{1}{2}\frac{\nabla^2\sqrt{\rho}}{\sqrt{\rho}} = & \ 0, \label{Bernoulli} \\
\partial_t\rho + \bm{\nabla}\cdot(\rho\bm{\nabla}\Phi) = & \ 0,\label{Continuity}
\end{align}
and $\mathbf{v}=\bm{\nabla}\Phi$ is identified as the velocity field of the flowing condensate. The first of these is similar to the Bernoulli equation from classical fluid dynamics, except for the final term which referred to as the quantum pressure. This term acts on scales on the order of the healing length. The second equation is simply a continuity equation for boson density. Together, \eqref{Bernoulli} and \eqref{Continuity} give a hydrodynamic interpretation to the flowing condensate.

\subsection{Vortex solutions}

Let us now specialise to a stationary condensate in polar coordinates $(r,\theta)$ with the wavefunction,
\begin{equation}
\Psi(r,\theta,t) = Y(r)e^{i\ell\theta-it},
\end{equation}
where the $e^{-it}$ dependence is the oscillation due to the chemical potential (which is unity in our dimensionless variables).
Since $\Psi(\theta=0)=\Psi(\theta=2\pi)$ by continuity of the wavefunction, the winding number $\ell$ is constrained to be an integer.
Assuming no flow in the radial direction (which amounts to taking $Y$ to be real) the velocity field is simply,
\begin{equation} \label{vortex}
\mathbf{v} = \frac{\ell}{r}\vec{\mathbf{e}}_\theta.
\end{equation}
This velocity profile is known as a vortex solution and, since $\ell$ can only take on integer values, the vortex circulation is said to be quantised.
In this work, we will take $\ell>0$ so that our vortex rotates in the direction of positive $\theta$, i.e. anti-clockwise.

\begin{figure} 
\centering
\includegraphics[width=\linewidth]{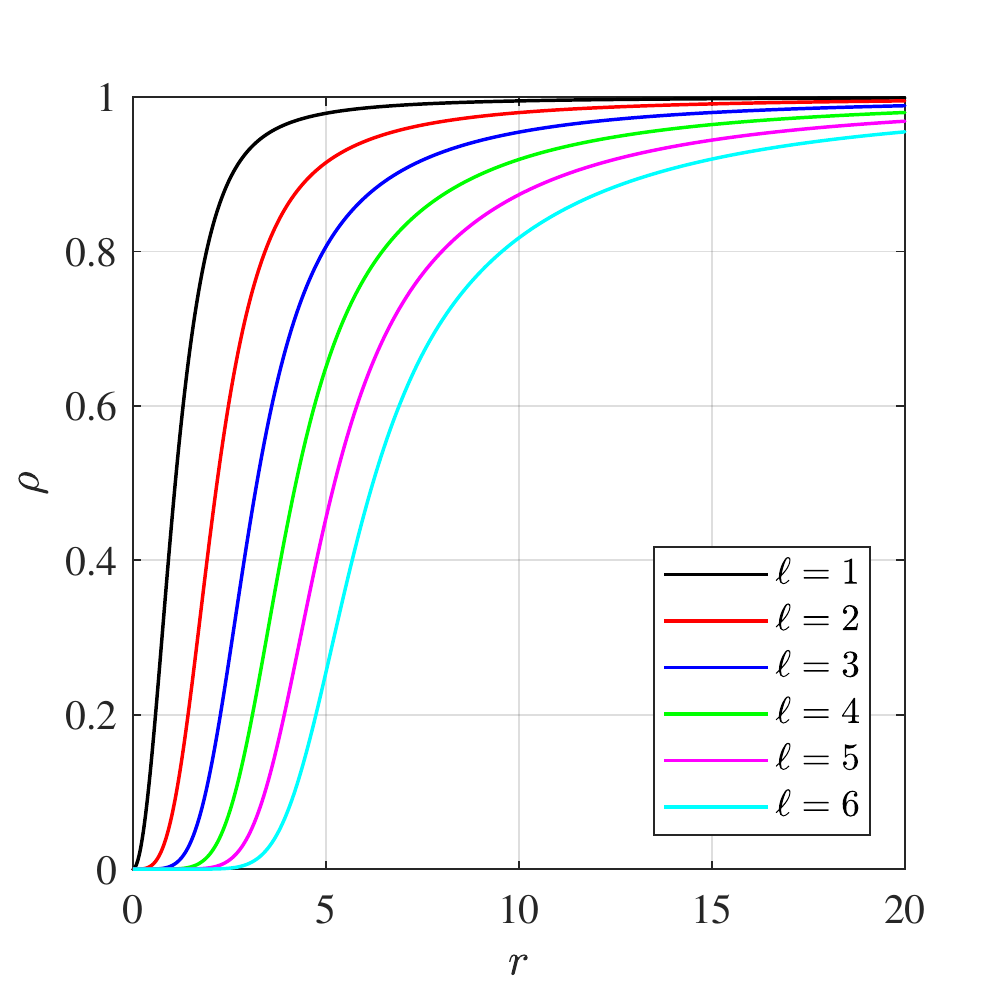}
\caption{The density profiles $\rho(r)$ for the first 6 non-zero $\ell$. It is apparent that, the larger the winding number $\ell$ is, the larger the vortex core is at a prescribed level of density.
} \label{fig:dens}
\end{figure}

The corresponding ground state density profile in an unbounded condensate, i.e. with $U=0$, is obtained by solving,
\begin{equation} \label{Yeq}
\frac{1}{2r}\partial_r\left(r\partial_r Y\right)+\left(1-\frac{\ell^2}{2r^2}\right)Y = Y^3,
\end{equation}
subject to the BCs $Y(r=0)=0$ and $Y(r\to\infty)=1$.
Then the density is obtained through $\rho=Y^2$ and the parameter $\rho_c$ introduced in \eqref{scales} is identified as the density of the vortex at spatial infinity.
In Fig.~\ref{fig:dens}, we show the numerically obtained density profiles corresponding to the first six values of $\ell$. The details of the numerical solver can be found in Appendix \ref{app:dens_vor}.
By noting that \eqref{Yeq} asymptotes to Bessel's equation in the limit $r\to 0$ and a polynomial equation for $r\to\infty$, we can write down the asymptotics for the density,
\begin{equation} \label{Z_asymps}
\rho(r)\sim\begin{cases}
\rho_0\,J^2_\ell(\sqrt{2}r), \qquad r\to 0, \\
1-\ell^2/2r^2, \ \qquad r\to\infty.
\end{cases}
\end{equation}
where $\rho_0$ is a constant amplitude. 

\subsection{Fluctuations}

Now we consider fluctuations of the condensate by writing $\Psi\to\Psi+\psi$ with $\psi\ll\Psi$.
Linearising the GPE \eqref{GPE} as well as its complex conjugate, the linear equations can be collected into a single matrix equation called the Bogoliubov-de Gennes (BdG) equation \cite{giacomelli2020ergoregion},
\begin{equation} \label{BdGeq}
i\partial_t|\psi\rangle = \widehat{L}|\psi\rangle,
\end{equation}
where $\psi$ has been split according to,
\begin{equation}
\begin{bmatrix}
\psi(t,r,\theta) \\ \psi^*(t,r,\theta)
\end{bmatrix} = \sum_{m=-\infty}^\infty e^{im\theta} \begin{bmatrix}
u_+(t,r)e^{+i\ell\theta-it} \\ u_-(t,r)e^{-i\ell\theta+it}
\end{bmatrix},
\end{equation}
and the operator $\widehat{L}$ and the state vector $|\psi\rangle$ are defined as,
\begin{equation}
\widehat{L} = \begin{bmatrix}
\mathcal{D}_+ & \rho \\ -\rho & -\mathcal{D}_-
\end{bmatrix}, \qquad |\psi\rangle = \begin{bmatrix}
u_+ \\ u_-
\end{bmatrix}
\end{equation}
with,
\begin{equation}
\mathcal{D}_\pm = -\frac{1}{2}\left[\partial_r^2+\frac{1}{r}\partial_r - \frac{(m\pm\ell)^2}{r^2}\right] + 2\rho + U -1.
\end{equation}
The BdG conserves the norm, which measures the particle content of a given mode and is defined as,
\begin{equation} \label{norm_tot}
    \mathcal{N} = \langle\psi|\sigma_3|\psi\rangle = \int d^2\mathbf{x}\left(|u_+|^2-|u_-|^2\right),
\end{equation}
where $\sigma_3$ is the third Pauli matrix.
$\partial_t\mathcal{N}=0$ then follows by inserting the BdG relation in \eqref{BdGeq} and invoking $u_\pm=0$ at either the system's outer boundary or as $r\to\infty$.

Alternatively, we can instead write down the linear equations in terms of fluctuations of the hydrodynamic variables defined in Section \ref{sec:hydro} by perturbing $\Phi\to\Phi+\phi$ and $\rho\to\rho\,(1+\eta)$. 
These variables are related to the perturbed wavefunction via,
\begin{equation} \label{psi}
\psi = \Psi\left(\tfrac{1}{2}\eta+i\phi\right),
\end{equation}
and obey the following set of coupled equations,
\begin{subequations} \label{lin}
\begin{align}
D_t\eta + \rho^{-1}\bm{\nabla}\cdot(\rho\bm{\nabla}\phi) & \ = 0, \label{lin1} \\
D_t\phi + \rho\eta - \tfrac{1}{4}\rho^{-1}\bm{\nabla}\cdot\left(\rho\bm{\nabla}\eta\right) & \ = 0, \label{lin2}
\end{align}
\end{subequations}
where $D_t = \partial_t+\mathbf{v}\cdot\bm{\nabla}$ is the material derivative.
Of course, \eqref{BdGeq} and \eqref{lin} are completely equivalent to one another, however, both representations will be used at different points in this work depending on which best suits our needs.

It is useful to separate the two real fields $\phi$ and $\eta$ into positive and negative frequency components,
\begin{subequations} \label{decomp}
\begin{align}
\phi = & \ \sum_\lambda\nolimits \left(\alpha_\lambda\varphi_\lambda+\alpha_\lambda^*\varphi_\lambda^*\right), \\ \eta = & \ \sum_\lambda\nolimits \left(\alpha_\lambda n_\lambda+\alpha_\lambda^*n^*_\lambda\right),
\end{align}
\end{subequations}
where the $\alpha_\lambda$ are constants and we have used the $t$ and $\theta$ independence of the background to write,
\begin{equation} \label{modes}
\begin{bmatrix}
\varphi_\lambda \\ n_\lambda
\end{bmatrix} = \begin{bmatrix}
\tilde{\varphi}(r) \\ \tilde{n}(r)
\end{bmatrix}e^{im\theta-i\omega t}.
\end{equation}
The sum over $\lambda$ is shorthand for,
\begin{equation}
\sum_\lambda\nolimits = \int^\infty_0 d\omega \sum_{m=-\infty}^{\infty}.
\end{equation}
Since the equations are linear by definition, each mode satisfies \eqref{lin} independently.
The mode equations can be derived from the following Lagrangian,
\begin{equation} \label{hydro_Lagr}
\begin{split}
\mathcal{L} = & \ \rho\bigg[\frac{1}{2}\left(\varphi_\lambda^* D_tn_\lambda + \varphi_\lambda	 D_tn^*_\lambda - n^*_\lambda D_t\varphi_\lambda - n_\lambda D_t\varphi^*_\lambda \right) \\
& \qquad -\rho\, n_\lambda n_\lambda^* - \bm{\nabla}\varphi_\lambda \cdot \bm{\nabla} \varphi^*_\lambda -\frac{1}{4}\bm{\nabla}n_\lambda \cdot \bm{\nabla}n^*_\lambda \bigg] \\
& \qquad -\frac{1}{2}\bm{\nabla}\cdot\left[\bm{\nabla}\rho \left(\varphi_\lambda\varphi_\lambda^*+\frac{1}{4}n_\lambda n_\lambda^*\right)\right].
\end{split}
\end{equation}
This is invariant under phase rotations, which results in a conservation equation for the norm $\mathcal{N}$, whose components are,
\begin{equation} \label{norm}
\begin{split}
\rho_n[\varphi] = & \ i\rho\left(\varphi_\lambda n_\lambda^*-\varphi_\lambda^* n_\lambda\right), \\
\mathbf{j}_n[\varphi] = & \ i\rho\big[\mathbf{v}\left(\varphi_\lambda n_\lambda^*-\varphi_\lambda^* n_\lambda\right) + \varphi_\lambda\bm{\nabla}\varphi_\lambda^* \\ & -\varphi_\lambda^*\bm{\nabla}\varphi_\lambda + \tfrac{1}{4}\left(n_\lambda\bm{\nabla}n_\lambda^*-n_\lambda^*\bm{\nabla}n_\lambda\right)\big],
\end{split}
\end{equation}
and satisfies,
\begin{equation} \label{current}
\partial_t\rho_n[\varphi]+\bm{\nabla}\cdot\mathbf{j}_n[\varphi] = 0.
\end{equation}
Using the relations $u_\pm = \rho^{-\frac{1}{2}}(\frac{1}{2}n_\lambda\pm i\varphi_\lambda)\alpha_\lambda e^{-i\omega t}$, one recovers the expression for the total norm in \eqref{norm_tot} by integrating the norm density over space, i.e. $\mathcal{N}=\int \rho_n\, d^2\mathbf{x}$.
Time translational invariance of \eqref{hydro_Lagr} gives rise to the conservation of mode energy. The conservation equation then contains the components $\rho_e$ and $\mathbf{j}_e$ where (for stationary set-ups) the relation to the components of the norm are $\rho_e=\omega\rho_n$ and $\mathbf{j}_e=\omega\mathbf{j}_n$. This reinforces the idea of the norm as the particle number, since the each particle in the mode contributes an energy proportional to $\omega$.
For our purposes, the norm density $\rho_n$  will be particularly useful in identifying the mechanism behind vortex instabilities and how they relate to superradiance.

\subsection{WKB method} \label{sec:wkb_method}

Since the equations in \eqref{lin} lack a closed form solution, we study them approximately using the WKB method.
The fundamental principle underlying the WKB method is the assumption that $\phi$ and $\eta$ look locally like plane waves.
That is, the amplitude of the waves varies slowly when compared with the phase or, equivalently, the background varies over much longer scales that the local wavelength.
In a homogeneous medium, plane waves are exact solutions and the WKB method therefore becomes exact.
The equations of motion then determine a relation between the wavevector $\mathbf{k}$ and the frequency $\omega$ of the waves called the dispersion relation.
In an inhomogeneous medium, the dispersion relation appears as the leading order solution in the WKB expansion and gives the relation between the local values of $\mathbf{k}$ and $\omega$.
Therefore, since the vortex profile in \eqref{vortex} depends only on the radial coordinate, the dispersion relation will determine the local value of the radial component of the wavevector for a given choice of $\omega$ and $m$.
As shown in e.g. \cite{patrick2020superradiance,patrick2021rotational}, one can then use the dispersion relation to study the scattering of the different modes in the system.
This will be the aim of the present section.

Before applying the WKB method, let us first write down the equations satisfied by the radial eigenfunctions $\tilde{\varphi}$ and $\tilde{n}$.
Plugging \eqref{modes} into \eqref{lin} results in the following coupled set of equations,
\begin{subequations} \label{rad}
\begin{align}
-i\Omega \tilde{n} + \frac{1}{\rho r}\partial_r\left(\rho r\partial_r\tilde{\varphi}\right) - \frac{m^2}{r^2}\tilde{\varphi} = & \ 0, \label{rad1} \\
-4i\Omega\tilde{\varphi} + 4\rho\tilde{n} - \frac{1}{\rho r}\partial_r\left(\rho r\partial_r\tilde{n}\right) + \frac{m^2}{r^2}\tilde{n} = & \ 0, \label{rad2}
\end{align}
\end{subequations}
where $\Omega$ is the frequency in the fluid frame,
\begin{equation} \label{Omega}
\Omega = \omega-\frac{m\ell}{r^2}.
\end{equation}
However, the naive application of the WKB approximation to the equations in \eqref{rad} does not yield the correct mode behaviour for the modes at the origin.
The simplest way to see what goes wrong is to define a new coordinate $y=\log r$ and rewrite \eqref{rad} as,
\begin{subequations} \label{rad3}
\begin{align}
\tilde{\varphi}'' + (\log\rho)'\tilde{\varphi}' - m^2\tilde{\varphi} = & \ i\Omega r^2 \tilde{n}, \\
\tilde{n}'' + (\log\rho)'\tilde{n}' - m^2\tilde{n} = & \ 4\rho r^2 \tilde{n} - 4i\Omega r^2 \tilde{\varphi},
\end{align}
where $'=\partial_y$.	
\end{subequations}
Then using the asymptotics of $\rho$, we recognise that $(\log\rho)'\to 2\ell$ approaching the origin which is comparable to $m$.
Since $m$ sets the scale on which the phase of the waves changes, the assumption of a separation of scales between the waves and the background (which is a requirement of WKB) is badly violated.
To proceed, we can factor out the offending term by writing,
\begin{equation} \label{new_modes}
\tilde{\varphi} = \rho^{-\frac{1}{2}}f, \qquad \tilde{n} = \rho^{-\frac{1}{2}}g,
\end{equation}
which reduces the equations in \eqref{rad3} to,
\begin{subequations} \label{new_eqs}
\begin{align}
f'' - \m^2 f = & \ i\Omega r^2 g, \\
g'' - \m^2 g = & \ 4\rho r^2 g -4i\Omega r^2 f,
\end{align}
\end{subequations}
where we have used the equation of motion for the density in \eqref{Yeq} to write,
\begin{equation} \label{mtild}
\m^2 = m^2 + \ell^2 + 2r^2(\rho-1).
\end{equation}
Since the factor of $\rho^{-\frac{1}{2}}$ in \eqref{new_modes} cancels that contained in $\Psi$ in \eqref{psi}, the new equations in \eqref{new_eqs} can be viewed as an intermediate step in converting the BdG equation \eqref{BdGeq} into the hydrodynamic equations \eqref{lin}.
The benefit of this rewriting is that there are no derivatives of any background quantities which become large approaching the origin.
Now we can write a WKB ansatz for the waves using the $y$ coordinate,
\begin{equation} \label{WKB_ansatz}
f(y) = \mathcal{A}(y)e^{i\int q(y) dy}, \quad g(y) = \mathcal{B}(y)e^{i\int q(y) dy},
\end{equation}
where $\mathcal{A}'\ll q\mathcal{A}$, $\mathcal{B}'\ll q\mathcal{B}$ and $q'\ll q^2$. 
Inserting into \eqref{new_eqs}, the leading order contributions can be combined into a single equation,
\begin{equation}
4\Omega^2 r^4 = 4\rho(q^2+\m^2) + (q^2+\m^2)^2.
\end{equation}
Then defining $q=pr$ and $k^2=p^2+\m^2/r^2$, this equation can be rewritten as a dispersion relation,
\begin{equation} \label{disp1}
\Omega^2 = F(k)k^2, \qquad F(k) = \rho + \tfrac{1}{4}k^2,
\end{equation}
which has two separate branches,
\begin{equation}
\omega_D^\pm = \frac{m\ell}{r^2}\pm\sqrt{F(k)k^2},
\end{equation}
where modes with $\omega=\omega_D^+$ are said to be on the upper branch and those with $\omega=\omega_D^-$ are on the lower branch.
We will shortly see that modes on the upper branch are the positive norm solutions whereas those on the lower branch have negative norm.

Since \eqref{disp1} is a quartic polynomial in $p$, it has four different solutions for $p$ which we call,
\begin{subequations}
\begin{align}
p^\pm(r) = & \ \pm\sqrt{-W_+}, \qquad p^{t,b}(r) = \mp i\sqrt{W_-}, \label{pmodes} \\
W_\pm = & \ \mp 2\sqrt{\rho^2+\Omega^2} +2\rho +\m^2/r^2,
\end{align}
\end{subequations}
The solutions $p^{t,b}$ are evanescent for all $r$ with $ip^t>0$ representing the mode which grows with increasing $r$ and $ip^b<0$ the one which decays.
The $p^\pm$ solutions correspond to in and out-going plane waves at large $r$.
The direction of propagation of these modes is determined by the radial component of the group velocity,
\begin{equation} \label{groupvel}
v^r_g = \partial_p\omega_D^\pm = \frac{\rho+k^2/2}{\Omega}p.
\end{equation}
Now that the phase has been determined, it remains to find an expression for the amplitudes.
In Appendix \ref{app:amp}, we show that the next to leading order equation gives an equation for the amplitude which can be put in the form,
\begin{equation} \label{WKB2amp}
\partial_r\left(rQ\mathcal{A}^2\right) = 0,
\end{equation}
with,
\begin{equation}
Q = F^{-1}\Omega v_g^r = \frac{\rho+k^2/2}{\rho+k^2/4}\,p,
\end{equation}
which we can solve for,
\begin{equation} \label{amp}
\mathcal{A} \propto \left|rQ\right|^{-\frac{1}{2}}.
\end{equation}
The amplitude $\mathcal{B}$ can then be found from the relation $F\mathcal{B}=i\Omega\mathcal{A}$, which is the leading order equation from \eqref{lin2}.
It is also useful to define two new amplitudes $A$ and $B$ which satisfy,
\begin{equation}
A = \sqrt{r}\mathcal{A}, \qquad B  = \sqrt{r}\mathcal{B},
\end{equation}
where the $\sqrt{r}$ factors out the part of the amplitude which grows when an in-going mode gets focussed onto a smaller disk.

Putting everything together, the original $\varphi_\lambda$ and $n_\lambda$ modes are given by,
\begin{subequations} \label{modez}
\begin{align}
\varphi_\lambda \sim & \ \left|r\rho F^{-1}\Omega v_g^r\right|^{-\frac{1}{2}}e^{i\int p(r)dr+im\theta-i\omega t}, \\
n_\lambda \sim & \ i\left|r\rho F\Omega^{-1}v_g^r\right|^{-\frac{1}{2}}e^{i\int p(r)dr+im\theta-i\omega t}.
\end{align}
\end{subequations}
Plugging these modes into the expression for the norm in \eqref{norm}, we find,
\begin{equation}
\rho_n[\varphi_\lambda] = 2\rho F^{-1}\Omega|\varphi_\lambda|^2.
\end{equation}
Recognising that $\Omega>0$ ($\Omega<0$) on the upper (lower) branch of the dispersion relation by definition, and also that $F$ is everywhere positive for propagating modes, we confirm the earlier statement that modes of the upper branch have positive norm whilst those on the lower branch have negative norm.
Thus, $\omega_D^\pm$ will henceforth be referred to as the positive/negative norm branches.
When a positive frequency mode is on the negative norm branch, the mode will have negative energy.
We will see later on (Sec.~\ref{sec:dbt_wkb}) how the region where this occurs can be naturally identified using our framework with the ergoregion, a notion commonly invoked BH physics \cite{brito2020superradiance}.

\subsection{Scattering} \label{sec:scatter}

Scattering occurs when one of the modes in \eqref{pmodes} gets converted into one of the others, which results in waves being (partially) reflected due to the inhomogeneity of the medium they propagate through.
The WKB approximation encodes no information about scattering, since it assumes that each mode evolves adiabatically as it moves through the flow.
Hence, the dominant contributions to scattering come from the locations in the flow where the WKB approximation fails.
An important example of where this occurs is at turning points in the radial direction, $r_\tp$, which are the locations where a wave radially stagnates, i.e. $v_g^r=0$, before reversing its direction.
From \eqref{groupvel}, we can see that this occurs when the local value of $p$ vanishes thereby causing the WKB amplitude to suddenly diverge, violating the slowly varying assumption.

From the expressions for the four $p$ in \eqref{pmodes}, we can see immediately that nowhere in the flow will $p^{t,b}=0$ be satisfied, since $W_->0$ everywhere. 
Hence, the $p^{t,b}$ modes do not scatter.
By contrast, since $W_+$ can be either positive or negative, it is possible to have $p^\pm=0$ and so these mode can scatter.
On one side of $r_\tp$, $W_+<0$ and the modes correspond to propagating solutions whereas on the other side, $W_+>0$ and the modes will be evanescent.
To avoid a naming ambiguity, let the $p^\pm$ be labelled $p^{\uparrow\downarrow}$ when they are evanescent: $p^\uparrow$ is the one which grows in the direction of increasing $r$ whereas $p^\downarrow$ is the one which decays.
The relation between the different mode amplitudes on either side of $r_\tp$ can be found from an asymptotic matching procedure, see e.g. \cite{patrick2020superradiance,patrick2020quasinormal} for relevant examples, and the results concisely expressed using the following transfer matrices,
\begin{equation} \label{transfer}
T = e^{\frac{i\pi}{4}}\begin{bmatrix}
1 & -\frac{i}{2} \\ -i & \frac{1}{2}
\end{bmatrix}, \qquad \widetilde{T} = e^{\frac{i\pi}{4}}\begin{bmatrix}
\frac{1}{2} & -\frac{i}{2} \\ -i & 1
\end{bmatrix}.
\end{equation}
These matrices are defined so that they act on the mode amplitudes at $r>r_\tp$ and return the amplitudes at $r<r_\tp$ according to the following relations,
\begin{equation} \label{transfer2}
\begin{bmatrix}
A^+_\tp \\ A^-_\tp
\end{bmatrix} = T \begin{bmatrix}
A^\downarrow_\tp \\ A^\uparrow_\tp
\end{bmatrix}, \qquad \begin{bmatrix}
A^\uparrow_\tp \\ A^\downarrow_\tp
\end{bmatrix} = \widetilde{T} \begin{bmatrix}
A^+_\tp \\ A^-_\tp
\end{bmatrix},
\end{equation}
where the labels $+,-,\uparrow,\downarrow$ correspond to the ones on the different $p$.
One can then construct a global solution to the equations of motion by using the WKB solutions in regions
of the flow separated by the $r_\tp$, then patching these solutions together using the transfer matrices.

\begin{figure*} 
\centering
\includegraphics[width=.8\linewidth]{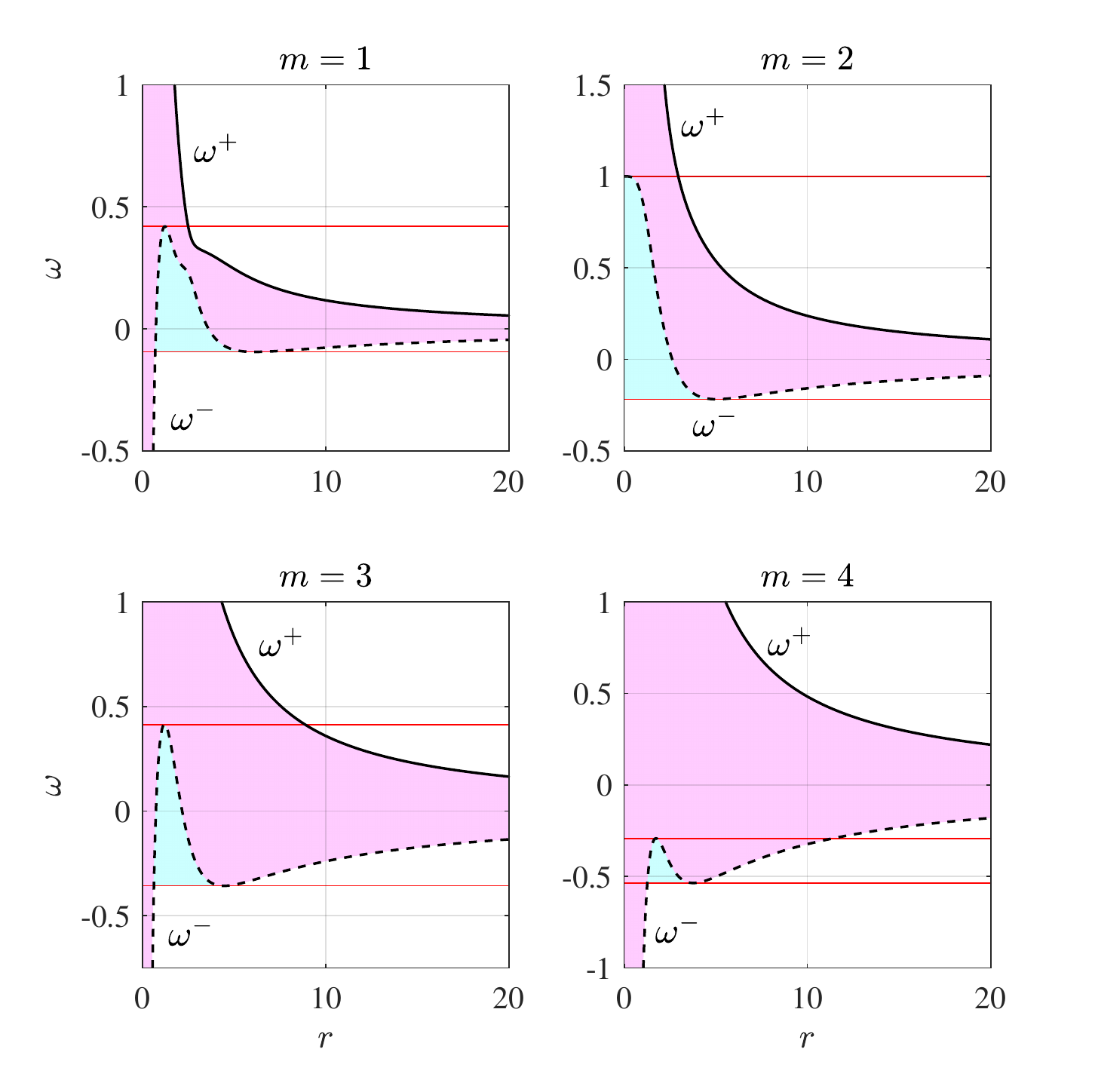}
\caption{The functions $\omega^+$ (solid black line) and $\omega^-$ (dashed black line) with $\ell=2$ for the first four $m>0$ values.
Modes above $\omega^+$ have positive norm whereas those below $\omega^-$ have negative norm.
Between the two curves (in the pink region) modes are evanescent, hence do not propagate.
In the frequency band between the horizontal red lines, waves can be trapped in a cavity in the vortex core (shaded in turquoise) and become (quasi-)bound states.
The possibility of there being modes with positive norm at infinity but negative norm in the cavity is what leads to the vortex instability.
Inverting the vertical axis about $\omega=0$ and swapping the labels $\omega^+\leftrightarrow\omega^-$ gives the corresponding plots for the $m<0$ modes.
} \label{fig:potentials}
\end{figure*}

There exists a useful representation that we can use to see where the turning points are located for different frequencies.
Plugging $p=0$ into \eqref{disp1} defines two curves,
\begin{equation} \label{potentials}
\omega^\pm(r) = \frac{m\ell}{r^2} \pm \sqrt{\rho\frac{\m^2}{r^2} + \frac{\m^4}{4r^4}},
\end{equation}
which give the frequency required for a mode to be on a turning point at a given radius.
Examples of these curves are displayed in Fig.~\ref{fig:potentials}.
Thus, we can find the locations of the turning points by looking for the intersections of a line with $\omega=\cst$ with the functions $\omega^\pm$.
In particular, when we have $\omega=\omega^+$ there will be a turning point on the positive norm branch of the dispersion relation, whereas for $\omega=\omega^-$ we have a turning point on the negative norm branch.
That means that if a particular frequency intersects with both the $\omega^\pm$ curves, there will be tunnelling between positive and negative norm branches of the dispersion relation.
This is the mechanism responsible for all the phenomena to be discussed in this work.

Lastly, let us note in passing that,
\begin{equation} \label{disp_potentials}
W_+W_- = -4(\omega-\omega^+)(\omega-\omega^-),
\end{equation}
which means that, since $W_->0$ everywhere, the zeros of $W_+$ coincide with the turning points, which can also be seen directly from \eqref{pmodes}.
Therefore one can think of $W_+$ as an effective potential governing the scattering of the $p^\pm$ modes.

\subsection{Mode asymptotics} \label{sec:asymp}

Having found a well-behaved set of WKB modes and looked at their scattering, it remains be to shown that these modes indeed incorporate the correct asymptotic behaviour.
Let us start with the limit that $r\to\infty$.
Using the limit for the density in \eqref{Z_asymps}, we find that $\m^2\to m^2$, see \eqref{mtild}.
Therefore the dispersion relation in \eqref{disp1} reduces to the well-known one for plane waves in uniform condensate (albeit in polar coordinates) as we would expect, see e.g. \cite{bogoliubov1947super,barcelo2011analogue}.

In the opposite limit $r\to 0$, we need to do a bit more work.
The first thing to notice is that when the density drops to zero, the BdG equation in \eqref{BdGeq} reduces to the usual Schr\"odinger equation for the decoupled $u_\pm$,
\begin{equation}
\frac{1}{2r}\partial_r\left(r\partial_ru_\pm\right) - \frac{(m\pm\ell)}{2r^2}u_\pm + (1\pm\omega)u_\pm = 0,
\end{equation}
which has the following general solution,
\begin{equation} \label{u_r0}
\begin{split}
u_\pm(r\to 0) = & \ \alpha_\pm J_{m\pm\ell}\big(\sqrt{2(1\pm\omega)}r\big) \\
& \qquad + \beta_\pm Y_{m\pm\ell}\big(\sqrt{2(1\pm\omega)}r\big),
\end{split}
\end{equation}
where $\alpha_\pm$ and $\beta_\pm$ are constants.
If our WKB modes have the correct limiting behaviour, they should agree with that in \eqref{u_r0} when taking $r\to 0$.
For this, we first need to know the asymptotic behaviour of the different $p$ in \eqref{pmodes}, which is found to be,
\begin{equation} \label{p_asymp}
p^2 \overset{r\to 0}{\sim} - \frac{(\ell\mp|m|)^2}{r^2} + 2\left[1\mp\omega\,\mathrm{sgn}(m)\right] + \mathcal{O}(r^2),
\end{equation}
with the upper sign corresponding to $p^\pm$ and the lower sign for $p^{t,b}$. The amplitudes are obtained by plugging the value of $p$ into \eqref{amp}.
Then to compare the WKB modes to the those in \eqref{u_r0}, we can use $u_\pm \sim\tfrac{1}{2}g\pm if$.
From \eqref{p_asymp}, it is readily apparent that there will be two distinctly different behaviours for $|m|\neq\ell$ and $|m|=\ell$, which we address now separately.

\subsubsection{Asymptotics for $|m|\neq\ell$}

In this case, the leading term in \eqref{p_asymp} as $r\to0$ is the one proportional to $r^{-2}$.
The phases of the different modes are,
\begin{equation}
e^{i\int p^{\uparrow\downarrow} dr} \sim r^{\pm||m|-\ell|}, \qquad e^{i\int p^{t,b} dr} \sim r^{\pm||m|+\ell|},
\end{equation}
and the amplitudes are constants to leading order.
We then recognise that the Bessel functions in \eqref{u_r0} behave as,
\begin{equation} \label{asymp_meql}
\begin{split}
J_{m+\ell}\sim r^{|m+\ell|}, & \qquad Y_{m+\ell}\sim r^{-|m+\ell|}, \\ J_{m-\ell}\sim r^{|m-\ell|}, & \qquad Y_{m-\ell}\sim r^{-|m-\ell|}.
\end{split}
\end{equation}
Hence, we find that the WKB modes do indeed incorporate the correct behaviour at the origin for $|m|\neq\ell$.

\subsubsection{Asymptotics for $|m|=\ell$} \label{sec:meql}

In this case, the asymptotic behaviour of the $p^{t,b}$ modes will be the same as the previous case with the $r^{-2}$ term dominating in \eqref{p_asymp}.
Correspondingly, either $u_+$ or $u_-$ (depending on the sign of $m$) in \eqref{u_r0} will match this behaviour.
For the $p^\pm$ modes, the leading term in \eqref{p_asymp} is the one which does not depend on $r$.
The behaviour of the modes now depends on both the sign of $m$ and the frequency.
For all $m<0$ modes and also for $m>0$ modes with $\omega<1$, the modes are oscillatory,
\begin{equation} \label{prop1}
e^{i\int p^\pm dr} \sim e^{\pm i|\varepsilon|r},
\end{equation}
whereas for $m>0$ with $\omega>1$ the modes are evanescent,
\begin{equation} \label{ev1}
e^{i\int p^{\uparrow\downarrow} dr} \sim e^{\pm|\varepsilon|r},
\end{equation}
where we have written $\varepsilon=\sqrt{2-2\,\mathrm{sgn}(m)\omega}$ for brevity.
The amplitudes in each of these cases scale with $r^{-\frac{1}{2}}$.
Now we compare to the solutions in \eqref{u_r0}.
The remaining solution, i.e. the one which does not match the $p^{t,b}$ modes, contain the functions,
\begin{equation} \label{asymp_mneql}
J_0\sim 1-\tfrac{1}{4}\varepsilon^2r^2, \qquad Y_0\sim\log r,
\end{equation}
which evidently do not match the behaviour of the WKB modes.
In fact, we could have already anticipated this since the WKB amplitude scaled with $r^{-\frac{1}{2}}$ which diverges at the origin, thereby violating the approximation.
This is not a serious problem since this is precisely the type of divergent behaviour that led to the definition of the transfer matrices around turning points in \eqref{transfer}.
Thus we can follow the same procedure here to remedy the situation.
To recap, the goal is to take the exact solution at the offending location (in this case a superposition of zeroth order Bessel functions), then match the asymptotics of these solutions onto the WKB modes.
For large arguments, the Bessel functions asymptote to,
\begin{equation}
\begin{split}
J_0(\varepsilon r) & \, \sim r^{-\frac{1}{2}}\cos\left(\varepsilon r-\pi/4\right), \\ Y_0(\varepsilon r) & \, \sim r^{-\frac{1}{2}}\sin\left(\varepsilon r-\pi/4\right).
\end{split}
\end{equation}
For the case of propagating modes at $r=0$ (which is the case we shall shortly be concerned with), the $J_0$ solution asymptotes to a combination of the modes in \eqref{prop1} with the amplitudes related by $A^-_0=iA^+_0$ whereas the combination of modes in $Y_0$ satisfy instead $A^-_0=-iA_0^+$.
Hence, although the WKB modes do not automatically contain the correct behaviour approaching the origin in this case, they can nonetheless be mapped onto the correct solutions using the same method we have already applied to deal with the turning points.

To summarize, using the WKB modes defined in \eqref{modez}, we can map our solution onto the correct asymptotic mode behaviour contained in the full equations of motion, which will ultimately allow us to implement the correct BCs.
This analysis also offers an intuitive perspective on the dispersion relation we are using in \eqref{disp1}, namely, it interpolates between the usual Bogoliubov-type dispersion for the excitations of a uniform condensate in limit $r\to\infty$ and the dispersion of the Schr\"odinger equation when the density drops to zero in the limit $r\to 0$.

\subsection{Open systems} \label{sec:vortex_open}

Having established the relevant preliminaries, we can now compute the instabilities of an $\ell$-charged vortex.
We consider first the case of an open system, i.e. one with no boundary at large $r$.

\subsubsection{Boundary conditions} \label{sec:open_BCs}

Since we have two coupled equations in \eqref{lin}, each of which being second order in spatial derivatives, we need to supply a total of four BCs.
Practically, this will allow us to eliminate all the mode amplitudes from the scattering computation, resulting in in a condition to be solved for the eigenfrequencies (as we will see in the next section).
The first BC is given by the fact that the natural modes of vibration in an open system are determined by the condition that there be no in-coming modes from infinity, i.e. $A^-_\infty=0$.
Furthermore, the amplitude of the $p^t$ mode diverges at infinity, so we need to take $A^t_\infty=0$.
Similarly, the amplitude of the $p^b$ mode di verges at the origin, hence we should also take $A^t_0=0$.
Since the $p^{t,b}$ modes do not scatter, these last two conditions fix the amplitudes of these modes to be zero everywhere, hence, we need not consider them any further in our calculation.

Before proceeding to state the final BC, consider the following.
If the system only contains either one or no turning points for a mode of a particular frequency, an in-coming mode will be completely reflected either at the turning point or at the origin.
We see then that setting the amplitude of the in-coming mode to zero trivially fixes the amplitude of the fourth (and final) mode in the system to also be zero.
Therefore the system has no natural vibrational modes in the frequency band containing containing one or no turning points.
What is needed then is a cavity capable of trapping modes such that the system can oscillate under no external influence.
Such a cavity exists when there are either two (for $|m|=\ell$) or three (for $|m|\neq\ell$) turning points, which can be seen, for example, in Fig.~\ref{fig:potentials}.

For $|m|\neq\ell$, let the turning points be labelled $r_0<r_1<r_2$.
Then, by using the $\widetilde{T}$ matrix in \eqref{transfer2}, we see that setting the amplitude of the mode which diverges at the origin to zero, i.e. $A^\downarrow_0=0$, gives a relation $A^-_0=iA^+_0$ between the propagating modes immediately outside of $r_0$.

Similarly, for $|m|=\ell$ we have the two turning points $r_1<r_2$ and also $r_0=0$.
From the analysis in Section~\ref{sec:meql}, we see that the combination of modes which yields a regular solution at $r=0$ is also $A^-_0=iA^+_0$.
Hence this labelling has allowed us to deal with the cases possessing both two and three turning points simultaneously.

In summary, the BCs for the excitations of a vortex in an open condensate are the following,
\begin{equation} \label{BCs1}
A^-_\infty=A^t_\infty=A^b_0 = 0, \qquad A^-_0=iA^+_0.
\end{equation}

\subsubsection{The resonance condition} \label{sec:res_cond}

To find the condition for the natural frequencies of the vortex, we first need to relate the amplitudes of the $p^\pm$ modes at $r_0$ to those as $r\to\infty$, which we achieve through,
\begin{widetext}
\begin{equation} \label{amp_matrix}
\begin{pmatrix}
A^+_0 \\ A^-_0 
\end{pmatrix} = \left|\frac{Q(\infty)}{Q(r_0)}\right|^\frac{1}{2}\begin{pmatrix}
e^{-iS_{01}} & 0 \\ 0 & e^{iS_{01}}
\end{pmatrix}T\begin{pmatrix}
0 & e^{S_{12}} \\ e^{-S_{12}} & 0
\end{pmatrix}\widetilde{T}\begin{pmatrix}
e^{-iS_{2\infty}} & 0 \\ 0 & e^{iS_{2\infty}}
\end{pmatrix}\begin{pmatrix}
A^+_\infty \\ A^-_\infty
\end{pmatrix},
\end{equation}
\end{widetext}
where,
\begin{equation}
S_{ij} = \int^{r_j}_{r_i}\left|W_+\right|^\frac{1}{2}dr,
\end{equation}
and we have used the fact that the factor $Q$ in \eqref{amp} is the same for both the $p^+$ and $p^-$ modes.
Implementing the BCs in \eqref{BCs1}, the matrix equation in \eqref{amp_matrix} can be solved for the following condition,
\begin{equation} \label{res_open}
e^{2iS_{01}} + X = 0, \qquad X = \frac{4+e^{-2S_{12}}}{4-e^{-2S_{12}}}.
\end{equation} 
For real frequencies, the first phase term has unit modulus whereas the second term is larger than 1.
Hence, \eqref{res_open} is only satisfied for complex frequencies $\omega_\mathbb{C}=\omega+i\Gamma$.
Indeed, these modes are quasi-bound states localised in the cavity $r\in[r_0,r_1]$ which gradually leak out to infinity by tunnelling through the barrier which exists in the region $r\in[r_1,r_2]$.
These modes can either grow or decay in time as we shall now see.

Let us assume that the characteristic growth/decay time is much longer than the period of oscillation of the modes, i.e. $|\Gamma|\ll|\omega|$, allowing us to perturb $S_{01}$ as,
\begin{equation}
S_{01}(\omega_\mathbb{C}) = S_{01}(\omega) + i\Gamma\partial_\omega S_{01}(\omega),
\end{equation} 
which is a good approximation provided the barrier trapping the modes is large (or $S_{12}\gg 1$).
To leading order in $\Gamma$, \eqref{res_open} splits into two separate conditions,
\begin{equation} \label{res_open2}
\cos S_{01}(\omega) = 0, \qquad \Gamma = -\frac{\log X(\omega)}{2\partial_\omega S_{01}(\omega)}.
\end{equation} 
The condition on the left in \eqref{res_open2} is the Bohr-Sommerfeld quantisation criterion which determines the real part of frequency.
This condition essentially states that a mode must have the correct wavelength to fit inside the cavity, with an additional phase shift coming from each of the turning points to account for the break down of WKB there.
The expression on the right of \eqref{res_open2} tells us the imaginary component of the frequency once the real part has been determined.
The $\log X$ term is always positive since $X>1$.
By contrast, the sign of the term $\partial_\omega S_{01}$ coincides with the sign of the norm of the modes trapped inside the cavity.
In other words, if a mode which solves \eqref{res_open2} is on $\omega_D^+$ everywhere then it has $\Gamma<0$ and the mode is stable, whilst if the mode tunnels from $\omega_D^+$ outside the cavity to $\omega_D^-$ inside of it, then it has $\Gamma>0$ and an instability occurs.
From the expressions for $\omega^\pm$ in \eqref{potentials}, tunnelling between the positive and negative norm branches of the dispersion relation  only occurs for $0<m<2\ell$, which can also be seen from Fig.~\ref{fig:potentials}.
Hence any instabilities that exist must lie in this range.

We comment in passing that there is an intuitive explanation for the form of $\Gamma$ in \eqref{res_open2}.
First, note that the time taken for the trapped mode to complete a full orbit inside $r\in[r_0,r_1]$ is given by $\tau = \pm2\partial_\omega S_{01}$, where the $\pm$ coincides with the sign of the norm of the mode in the cavity.
Rearranging, we find that $\exp(\Gamma\tau) = |\mathcal{R}|$, where $|\mathcal{R}|=X^{\mp 1}$ is the modulus of the reflection coefficient associated with the barrier.
Hence, the amount of growth/decay that occurs over a complete orbit of a wave in the cavity is determined by the size of the wave reflected back into it.

\subsubsection{Computing the instabilities}

\begin{figure} 
\centering
\includegraphics[width=\linewidth]{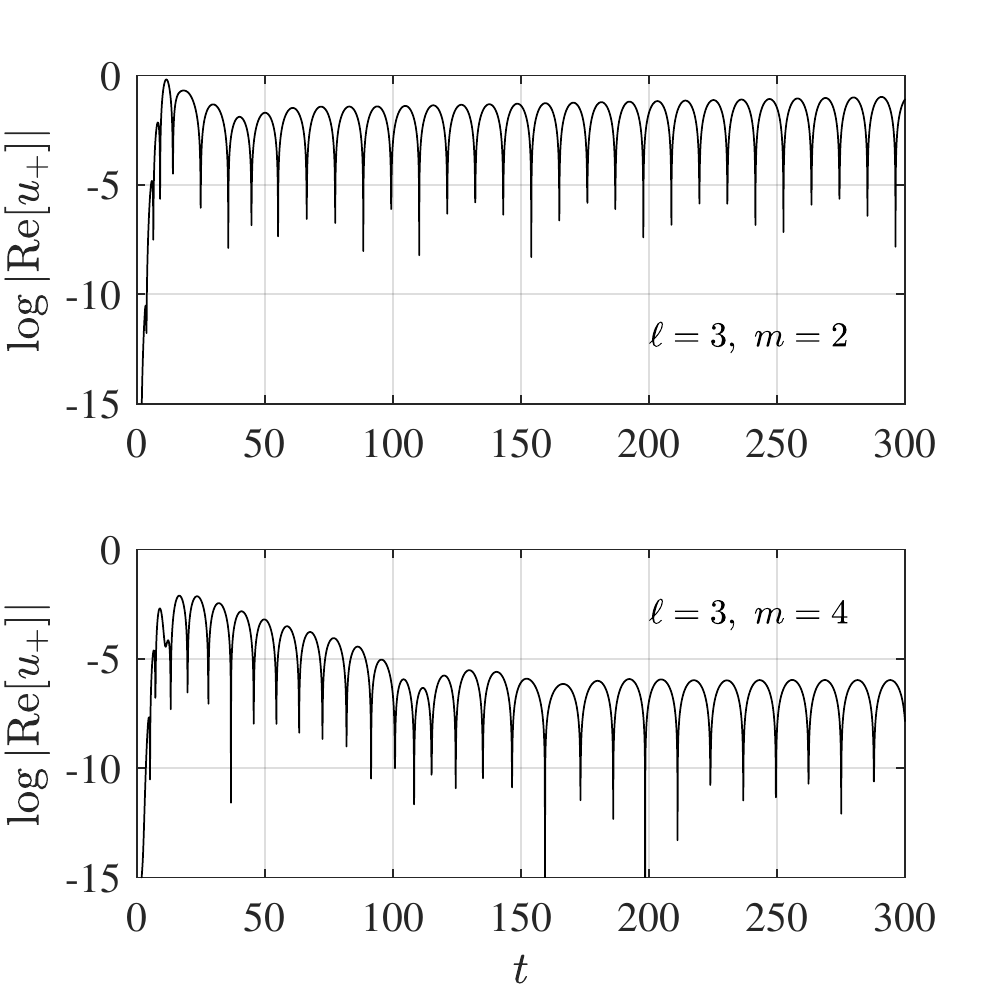}
\caption{Examples of the vortex response in the time domain following stimulation by a Gaussian pulse.
In the upper panel, only the instability (the growing mode) is present.
The lower panel gives an example of a stable mode which is present at early times and an unstable mode at late times.
If the magnitude of the growth/decay rate of these different modes is comparable in magnitude, it is not possible to extract the imaginary component of the frequency.
Furthermore, if the growth of the mode is too slow, it is not possible to measure over the window of observation.
} \label{fig:RK4_sims}
\end{figure}

We compute the unstable complex frequencies with $\omega>0$ by solving the approximate resonance condition in \eqref{res_open2}.
In Table~\ref{tab:instabs}, we display all unstable modes that occur for $|m|\leq 4$ and $\ell\leq 4$.
We find that the $\ell=1$ vortex is stable and that the $m=2$ mode is the most unstable for all the other $\ell$ probed.
This makes sense in light of the scattering picture in Fig.~\ref{fig:potentials} since the barrier is smallest for low $m$.
We also see that, at fixed $m$ and $\ell$, the modes with the largest frequency will be the fastest to grow, since for small frequencies the barrier is larger which inhibits the growth of the mode.
As $\ell$ and $m$ increase, the size of the cavity increases such that it becomes possible to fit more than one instability inside, as in the case of $m=\ell=4$.

We cross validate our WKB results with a time domain simulation of the BdG equation in \eqref{BdGeq}.
The equations are evolved forward in time via a trapezoidal method with 5-point centred finite difference stencils for the spatial derivatives (see Appendix \ref{app:BdG_sims} for more details).
Using the behaviour of the modes at the origin as discussed in Section \ref{sec:asymp}, we apply a Dirichlet BC at $r=0$ for $|m|\neq\ell$ and a Neumann BC for $|m|=\ell$.
We apply a Dirichlet BC at the outer boundary, which is placed sufficiently far away that reflections from the boundary do not enter our window of analysis.
For each $\ell$ and $m$, we send a Gaussian pulse in towards the centre of the vortex and after the initial pulse has passed, we are able to measure the response of the vortex.
Examples of the vortex response can be seen in Fig.~\ref{fig:RK4_sims}.

The oscillation frequency is extracted by performing a time Fourier transform of the signal and finding the frequency with maximum amplitude.
The growth rate is obtained by fitting the envelope of the signal with an exponential function where possible.
The results are displayed in Table~\ref{tab:instabs} and display excellent agreement in the real part of the frequency with those found using \eqref{res_open2}.
In particular, the agreement improves for larger $\omega$, as one would expect of the WKB approximation.
This confirms that our method can be used to provide insight into the mechanism underpinning the instability.
In all cases except for the $m=2$ mode, we were unable to extract the value of $\Gamma$, either because the growth rate was too slow to be resolved within our window of observation or due to interference between multiple modes in the system; namely between different instabilities, e.g. for $m=\ell=4$ or between the unstable mode of interest and a decaying mode (not quoted in Table~\ref{tab:instabs}) which we are not concerned with here.

 \begin{table}
 \centering
 \begin{tabular}{|c|l|r|r|r|r|}
 \hline
 $\ell$ & \multicolumn{1}{c|}{$m$} & \multicolumn{1}{c|}{$\omega_\mathrm{WKB}$} & \multicolumn{1}{c|}{$\omega_\mathrm{sim}$} & \multicolumn{1}{c|}{$\Gamma_\mathrm{WKB}$} & \multicolumn{1}{c|}{$\Gamma_\mathrm{sim}$} \\
 \hline
 
 \parbox[t]{2mm}{\multirow{1}{*}{$2$}} 
 & $2$ & $0.3954$ & $0.4399$ & $2.543\times 10^{-3}$ & $2.431\times 10^{-3}$ \\
 \hline
 
 \parbox[t]{2mm}{\multirow{3}{*}{$3$}}
 & $2$ & $0.2475$ & $0.2871$ & $1.903\times 10^{-3}$ & $2.189\times 10^{-3}$ \\
 & $3$ & $0.6613$ & $0.6648$ & $7.839\times 10^{-4}$ & \textit{n.a.} \\
 & $4$ & $0.2161$ & $0.2462$ & $2.370\times 10^{-7}$ & \textit{n.a.} \\
 \hline
 
 \parbox[t]{2mm}{\multirow{4}{*}{$4$}} 
 & $2$ & $0.1851$ & $0.2138$ & $1.593\times 10^{-3}$ & $1.832\times 10^{-3}$ \\
 & $3$ & $0.4773$ & $0.4995$ & $8.667\times 10^{-4}$ & \textit{n.a.} \\
 & $4$ & $0.1163$ & $0.1324$ & $3.406\times 10^{-8}$ & \textit{n.a.} \\
 &  & $0.7908$ & $0.7804$ & $1.711\times 10^{-4}$ & \textit{n.a.} \\
 \hline
 \end{tabular}
 \caption{Unstable frequencies $\omega+i\Gamma$ of a vortex in an open system are computed using the WKB result in \eqref{res_open2} and by a time domain simulation of \eqref{BdGeq}. We display all the unstable frequencies for $|\ell|\leq 4$ and $|m|\leq 4$, and quote the values to four digit precision. At given $\ell$ and $m$, modes with lower $\omega$ are deeper in the cavity and thus have a slower growth rate, leading to difficulties in extracting $\Gamma_\mathrm{sim}$ from simulations. If the growth rate could not be reliably extracted, we display \textit{n.a.} in the last column.
} \label{tab:instabs}
 \end{table}

\subsection{Closed systems} \label{sec:vortex_close}

Truly open systems are problematic to probe both experimentally due to the difficulty in imposing purely out-going BCs at the edge of the system.
Indeed, it is much more natural to impose a trapping potential $U$ which confines the condensate to a spatially finite domain.
The consequence of this is that any out-going modes will be reflected back into the condensate, causing the resonance condition to be modified.
Instead of the quasi-bound states of the previous section, the system will then exhibit true bound states which are the normal modes of the condensate.
However, it is still possible to have instabilities in the spectrum as we will now show.
A detailed study of the real time dynamics of vortex decay as obtained by the GPE evolution can be found in our companion paper \cite{patrick2021origin}.

\subsubsection{Boundary conditions}

\begin{figure} 
\centering
\includegraphics[width=\linewidth]{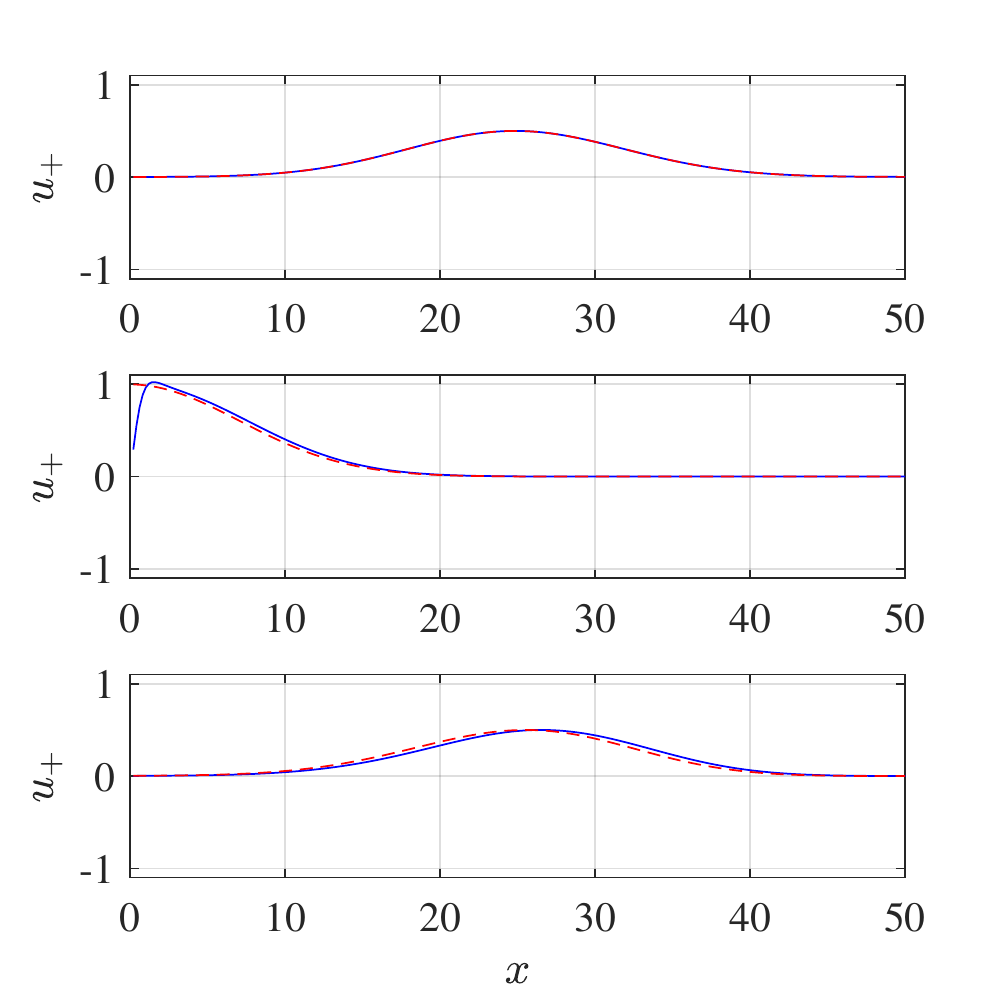}
\caption{Initially left moving pulse, whose evolution is dictated by 1D BdG, colliding with a hard wall at $x=0$. Three different times are shown, $t=0$ (top) $t=1250$ (middle) and $t=2500$ (bottom).
The blue curve uses the true BC at $x=0$ for $u_+$ (Dirichlet) with the background density profile $\rho=\tanh^2x$.
We find good agreement with the red dashed curved, which uses a Neumann BC with $\rho=1$.
} \label{fig:1D_refl}
\end{figure}

Let us approximate the potential as an infinite barrier located at $r_B$, that it $U(r<r_B)=0$ and $U(r_B)=\infty$, with $r_B$ a large but finite radius.
Since the condensate wavefunction vanishes for $r>r_B$, we must have $\Psi(r=r_B)=0$.
Taking $r_B\gg\ell$, the equation of motion for the condensate wavefunction in \eqref{Yeq} adopts the form,
\begin{equation}
\partial_r^2 Y + 2Y(1-Y^2) \simeq 0,
\end{equation}
near the boundary.
The solution to this equation is,
\begin{equation}
Y\approx \tanh(r_B-r),
\end{equation}
which drops quickly from the bulk value of $\approx 1$ to zero within a few healing lengths (due to the quantum pressure term in \eqref{Bernoulli} \cite{pethick2008bose}).
The question of determining the correct BCs at $r_B$ then becomes equivalent to asking how plane waves in a 1D condensate are reflected by a hard wall.
In Fig.~\ref{fig:1D_refl}, we verify that a pulse incident on a hard wall (comprised of wavelengths larger than the healing length) reflects analogously to waves propagating through a homogeneous condensate with a Neumann BC at $r_B$, that is $A^-_B=A^+_B$.
Since the instability is a low frequency phenomenon, we will therefore assume in our WKB analysis that the density has the same form of that of the free vortex all the way up to $r_B$ where we impose $A^-_B=A_B^+$ on the propagating modes.

The remaining BCs are imposed by noting that the behaviour of the modes at the origin is the same as in an open system, therefore at $r=0$ we have the same BCs as in the case of the free vortex.
Finally, since the $p^{t,b}$ mode will also be completely reflected at $r_B$ and these modes do not scatter anywhere else in the system, their amplitude must be zero everywhere since the $p^b$ mode diverges at the origin.

In summary, the BCs for the excitations of a vortex in a bounded condensate with a hard wall located at $r_B$ are,
\begin{equation} \label{BCs2}
A^t_B=A^b_0 = 0, \qquad A^-_0=iA^+_0, \qquad A^-_B=A^+_B.
\end{equation}

\subsubsection{The resonance condition}

The normal mode spectrum of the vortex can now be obtained from a resonance condition similar to the one derived in Section~\ref{sec:res_cond}.
We focus our attention on scenarios where there are, again, either two or three turning points using the labelling convention for $r_0<r_1<r_2$ already established.
Note that in a closed system, there will also be normal modes in the frequency bands containing either one or no turning points due to the box-like BCs, however, these bands do not contain instabilities which are our main interest.
The matrix equation relating the amplitudes at $r_0$ and $r_B$ is the same as in \eqref{amp_matrix} (provided we interpret the amplitudes on the right of the expression as the ones at $r_B$).

\begin{figure} 
\centering
\includegraphics[width=\linewidth]{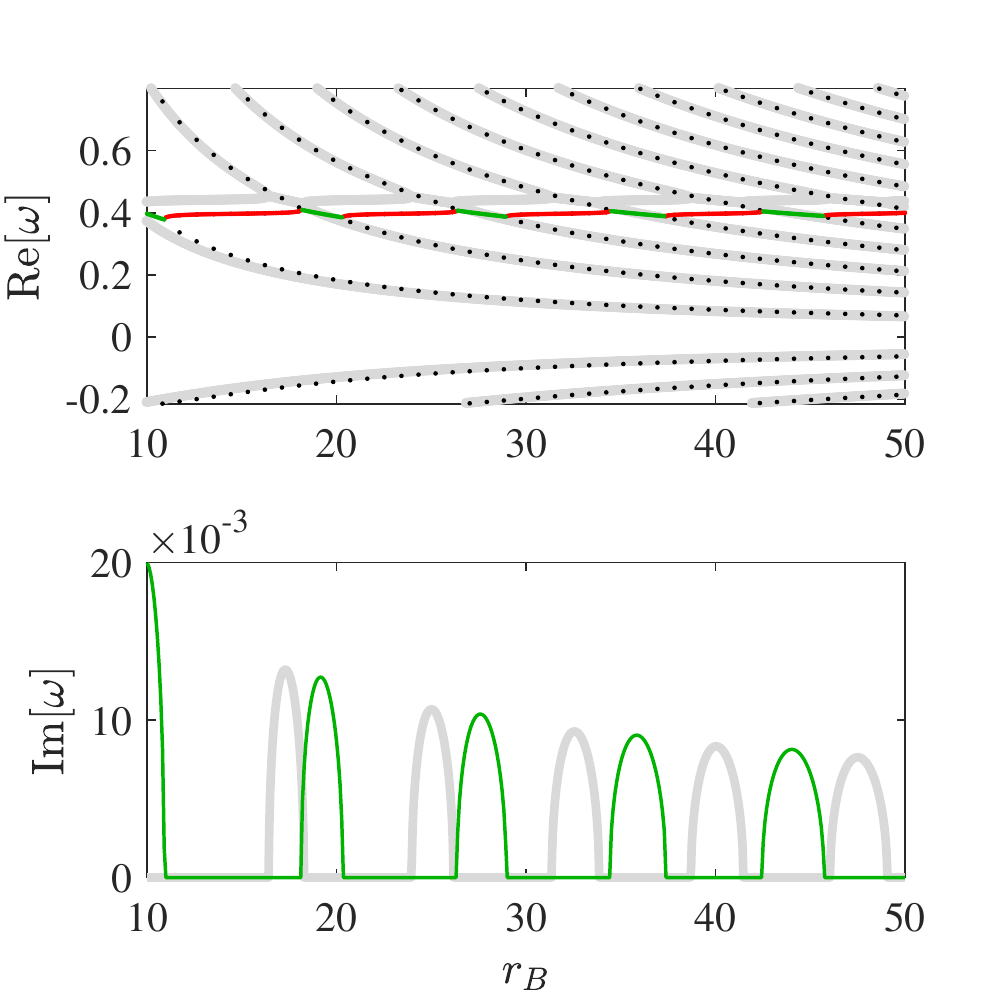}
\caption{The spectrum of the vortex for $m=\ell=2$ as a function of the system size $r_B$.
The upper panel gives the oscillation frequency. Modes localised inside (outside) the cavity are shown in black (red).
Note that the total energy (norm multiplied by frequency) of the modes multiplied by the frequency is positive for black and negative for red.
When these two types of modes cross each other, the resulting modes have zero total norm (green) \cite{giacomelli2020ergoregion} and are unstable.
The imaginary part of frequency of these unstable modes is shown in the lower panel.
This plot is a stability map, illustrating that the instability occurs in bubbles of finite width.
Eigenfrequencies computed from a numerical solution of \eqref{BdGeq} are shown in grey for comparison.
} \label{fig:stab_map}
\end{figure}

Applying the BCs in \eqref{BCs2}, the resulting condition can be cast into the following form,
\begin{equation} \label{res_clos1}
4\cot(S_{01})\cot(S_{2B}+\pi/4) = e^{-2S_{12}}.
\end{equation}
We solve this condition for the case of $m=\ell=2$ for a variety of $r_B$ and display the spectrum of modes in Fig.~\ref{fig:stab_map}.

Normal modes which are localised inside the cavity $r\in[r_0,r_1]$ are shown in red, whereas the black points are modes which are live outside in the region $r\in[r_2,r_B]$.
The green points represent a resonant coupling between these two different types of modes which leads to an instability (as we will soon discuss).
Also on Fig.~\ref{fig:stab_map}, we display (in grey) the eigenfrequencies as determined by a numerical resolution of the BdG equation in \eqref{BdGeq}. 
These are found as the eigenvalues of the BdG operator $\widehat{L}$, where the radial derivatives in $\widehat{L}$ are expressed using 5-point centred finite different stencils which encode the boundary conditions.
The WKB method agrees exceptionally well with the exact solution for modes localised outside the cavity since the flow is slowly varying in this region.
The agreement for modes localised inside the cavity, although still satisfactory, is reduced since the flow varies more rapidly in this region (see e.g. the density profiles in Fig.~\ref{fig:dens}).
Crucially, however, the WKB spectrum captures all the essential features of the exact spectrum (in particular the crossing of positive and negative norm modes) meaning we can use our WKB formula in \eqref{res_clos1} to interpret the instability mechanism.

The features of the plots in Fig.~\ref{fig:stab_map} can be understood purely from the form of the condition in \eqref{res_clos1}.
If the zeros of the two cotangent functions are well separated, the resonance condition is satisfied when one of these is exponentially small (which we approximate as zero).
Therefore, we either have $\cot(S_{01})=0$, which gives the red data in Fig.~\ref{fig:stab_map}, or $\cot(S_{2B}+\pi/4)=0$, which gives the black dots.
However, if the zeros of the cotangents are close together, it is no longer correct to approximate $\cot\simeq0$ since only the product of the two needs to be exponentially small.
This leads to a deviation in behaviour away from the curves traced out by the red and black data, which is precisely where the green dots appear.

When this coupling between the two types of modes occurs, the mode outside the vortex transmits negative norm into the cavity when it scatters with the barrier, leading to amplification of the mode outside the cavity.
Similarly, the mode inside the cavity transmits more positive norm to the outside of the cavity as it reflects off the barrier, leading to amplification inside.
This continual amplification of the mode causes it to become unstable.
Note that during this process, norm conservation is never broken since positive and negative norm are created in equal amounts.
Indeed, an identity resulting from the BdG equation \eqref{BdGeq} is,
\begin{equation}
(\omega-\omega^*)\int d^2\mathbf{x} \left(|u_+|^2-|u_-|^2\right) = 0,
\end{equation}
where the integral is the total norm from \eqref{norm_tot}.
This shows that for unstable modes with $\mathrm{Im}[\omega]\neq0$, the eigenmodes have zero total norm.

It has also been noted in the literature \cite{takeuchi2018doubly,giacomelli2020ergoregion} that in the limit that the system becomes infinitely large, i.e. $r_B\to\infty$, the frequency of the instability should asymptote to that in the open system.
This is somewhat counter intuitive since the eigenfrequencies of a system are highly dependent on the BCs, and imposing open/closed BCs are completely different procedures.
Our condition in \eqref{res_clos1} offers some insight into this conundrum. 
Rearranging, we find,
\begin{equation}
\begin{split}
e^{2iS_{01}} + X = & \ i e^{2iS_{2B}}\left(Xe^{2iS_{2B}}+1\right), \\
\sim & \ \mathcal{O}\left(e^{-2\Gamma\tau_B}\right),
\end{split}
\end{equation}
where $\tau_\infty = \partial_\omega S_{2B}$ is the time taken for the mode to propagate from just outside the barrier to the outer boundary, and we have assumed a slow growth rate with $|\Gamma|\ll|\omega|$.
Hence, as $r_B\to\infty$ we also have $\tau_B\to\infty$ and the resonance condition for the closed system reduces to that of the open one in \eqref{res_open}.
Physically, this is due to the fact that the instability originates inside the cavity, so its amplitude at small $r$ will be exponentially larger than the mode reflected by the boundary at large $r$. 
In the limit where the boundary is infinitely far away, the amplitude of the reflected mode will be infinitely smaller than the out-going mode and can therefore be discarded.

Finally, this resonant coupling picture also provides a useful perspective on the vortex instability in the open system.
In the vortex core, the normal modes are those which roughly fit in the cavity whereas outside the vortex, the spectrum of normal modes is continuous since the system in that case is infinite.
Therefore, the existence of the instability in the open system depends only on the existence of a trapped mode in the cavity, since there is guaranteed to be a mode outside which has the right frequency for a coupling to occur.

\section{Black holes}

In this section, we will study superradiance around  BH spacetimes and illustrate the connection with the vortex instabilities of the previous section.
Before looking at a gravitational (Kerr) BH, we first consider an analogue BH in order to bridge the gap with the vortices we have already looked at.

\subsection{Draining vortex} \label{sec:dbt_vortex}

Consider a BEC containing a vortex of the type \eqref{vortex}, but now with a draining component in the velocity field.
It is well-known that the resulting system shares many features with a rotating BH spacetime.
In the region where the density is approximately uniform, i.e. $\rho=1$, the continuity equation \eqref{Continuity} gives the following solution for the velocity field,
\begin{equation} \label{drain_vel}
\mathbf{v} = -\frac{d}{r}\,\vec{\mathbf{e}}_r + \frac{\ell}{r}\,\vec{\mathbf{e}}_\theta,
\end{equation}
where we have adopted the units established in Section~\ref{sec:GPE} and we take $d>0$ since the system is draining.
To establish such a set up, one can imagine pumping atoms out of the condensate at a rate $\dot{N}=-d$ in a small region around the vortex axis $r=0$.
Note that \eqref{drain_vel} is only an estimate and the full fluid flow is more complicated in both classical \cite{andersen2003anatomy,stepanyants2008stationary} and quantum \cite{zezyulin2014stationary} fluids. However, this approximation is known to hold far from the vortex axis, e.g. \cite{torres2017rotational}, and as such, can be treated as the leading order solution in a large $r$ expansion.

Excitations of the flow satisfy the equations of motion in \eqref{lin}.
Assuming that $d\gg 1$, so that the characteristic draining length scale is much larger than the healing length, the relevant frequencies to the problem will be smaller than $\mathcal{O}(1)$ and the final term in \eqref{lin2} (which arises from the quantum pressure) can be neglected. The two equations in \eqref{lin} now combine into a single wave equation,
\begin{equation} \label{wave_eqn}
D_t^2\phi - \nabla^2\phi = 0.
\end{equation}
This wave equation describes long wavelength phonons propagating at speed $c=\sqrt{\mu/M}$ (which is 1 in these units) independent of their frequency.
This approximation therefore neglects the effects of dispersion.

The wave equation in \eqref{wave_eqn} is formally equivalent to the equation of motion for a massless Klein-Gordon (KG) field propagating through a curved spacetime,
\begin{equation}
\frac{1}{\sqrt{-g}}\partial_\mu\left(\sqrt{-g}g^{\mu\nu}\partial_\nu\phi\right) = 0,
\end{equation}
where the effective metric entering this equation is given by,
\begin{equation} \label{DBT_metric}
\begin{split}
g^{\mu\nu}dx^\mu dx^\nu = & \ -\left(1-\frac{\ell^2+d^2}{r^2}\right)dt^2 + \frac{2d}{r}dr\,dt \\
& \qquad - 2\ell\,d\theta\,dt + dr^2 + r^2d\theta^2,
\end{split}
\end{equation}
with coordinates $(t,r,\theta)$.
Here we take the metric signature as $(-1,1,1,1)$ and $g$ is the determinant of the metric.
The metric \eqref{DBT_metric} can be written in a form where the BH behaviour is more apparent by performing a coordinate transformation $(t,\theta)\to(\tilde{t},\tilde{\theta})$ with,
\begin{equation}
d\tilde{t} = dt + \frac{v_r dr}{1-v_r^2}, \qquad r d\tilde{\theta} = r d\theta + \frac{v_\theta v_r dr}{1-v_r^2},
\end{equation}
where we have written $v_i = \vec{\mathbf{e}}_i\cdot\mathbf{v}$.
The metric becomes,
\begin{equation}
\begin{split}
\tilde{g}_{\mu\nu} d\tilde{x}^\mu d\tilde{x}^\nu = & \ -\left(1-\frac{r_e^2}{r^2}\right)d\tilde{t}^{\,2} + \frac{dr}{1-r_h^2/r^2} \\
& \qquad - 2\ell\,d\tilde{\theta}\,d\tilde{t} + r^2d\tilde{\theta}^2,
\end{split}
\end{equation}
with $\tilde{x}=(\tilde{t},\tilde{\theta},r)$ and we have defined,
\begin{equation} \label{re_rh}
r_e = \sqrt{\ell^2+d^2}, \qquad r_h = d,
\end{equation}
which satisfy $\tilde{g}_{\tilde{t}\tilde{t}}(r_e)=0$ and $\tilde{g}_{rr}(r_h)\to\infty$.
These locations are ergosphere $r_e$, the boundary of the region in which a phonon will be forced to co-rotate with the vortex as compared to a static point at infinity (i.e. the ergoregion), and the horizon $r_h$, the radius inside which all phonons are forced to move inward (i.e. toward $r=0$).
This is markedly different from the non-draining vortex discussed earlier, which had an ergoregion but no horizon. We will shortly use the WKB method to make the connection between $r_e$ (defined using the metric above) and the ergoregion of the non-draining vortex (which we defined as the region where excitations could have negative energy).

\subsubsection{Conventional method} \label{sec:SR_con}

Here we recap the conventional analysis used to study BH superradiance, which has been applied to this particular draining vortex set-up on multiple occasions. More details can be found in e.g. \cite{basak2003superresonance,basak2003reflection,berti2004qnm,basak2005analog,slatyer2005superradiant,slatyer2005superradiant,federici2006superradiance,richartz2015rotating,churilov2018scattering,demirkaya2020acoustic}.

To begin, we split the field into the different $\omega$ and $m$ modes,
\begin{equation} \label{decomp1}
\begin{split}
\phi = & \ \frac{R(r)}{\sqrt{r}}e^{im\tilde{\theta}-i\omega\tilde{t}}, \\
= & \ \frac{R(r)}{\sqrt{r}}\exp\left(-i\int\frac{v_r\tilde{\omega}\,dr}{1-v_r^2}\right)e^{im\theta-i\omega t},
\end{split}
\end{equation}
with a generic wave consisting of a sum of these modes, as in \eqref{decomp}, and we have defined,
\begin{equation} \label{om_tild}
\tilde{\omega} = \omega - \frac{m\ell}{r^2}.
\end{equation}
Next, one introduces a new coordinate (which in GR is called the tortoise coordinate),
\begin{equation} \label{tortoise}
r_* = r + r_h\log\left|\frac{r-r_h}{r+r_h}\right|^\frac{1}{2},
\end{equation}
thereby mapping $r\in(r_h,\infty)$ onto $r_*\in(-\infty,\infty)$.
Plugging \eqref{decomp1} and \eqref{tortoise} into \eqref{wave_eqn}, the problem of a wave scattering with a BH is converted into 1D scattering problem,
\begin{equation} \label{1dscat}
-\partial_{r_*}^2 R + V R = 0,
\end{equation}
in the presence of an effective potential barrier,
\begin{equation} \label{dbt_pot}
\begin{split}
V = & \ -\left(\omega-\frac{m\ell}{r^2}\right)^2 + \left(1-\frac{d^2}{r^2}\right) \\
& \qquad \qquad \qquad \qquad \times\left(\frac{m^2-\frac{1}{4}}{r^2}+\frac{5d^2}{4r^4}\right).
\end{split}
\end{equation}
The scattering equation in \eqref{1dscat} has the asymptotic solutions,
\begin{equation} \label{asymp_dbt}
R \sim \begin{cases}
A_h e^{-i\tilde{\omega}_hr_*}, \,\, \qquad \qquad \qquad r_*\to -\infty \\ 
A^-_\infty e^{-i\omega r_*} + A^+_\infty e^{i\omega r_*} \qquad r_*\to +\infty
\end{cases}
\end{equation}
where $\tilde{\omega}_h=\omega-m\Omega_h$ is the quantity in \eqref{om_tild} evaluated on the horizon.
Furthermore, an equation of the form \eqref{1dscat} admits a conserved quantity called the Wronskian,
\begin{equation} \label{wronsk}
\partial_{r_*}\left(R^*\partial_{r_*}R-R\,\partial_{r_*}R^*\right)=0,
\end{equation}
which is equivalent to the conserved norm in \eqref{norm}.
Finally, plugging the solutions \eqref{asymp_dbt} in \eqref{wronsk} one finds that the relation,
\begin{equation}
|\mathcal{R}|^2+\frac{\tilde{\omega}_h}{\omega}|\mathcal{T}|^2 = 1,
\end{equation}
is satisfied, where the reflection $\mathcal{R}$ and transmission $\mathcal{T}$ coeffcients are defined,
\begin{equation} \label{scat_coefs}
\mathcal{R}=A^+_\infty/A^-_\infty, \qquad \mathcal{T}=A_h/A^-_\infty.
\end{equation}
Thus, one deduces that amplification of the incident wave, i.e. $|\mathcal{R}|>1$, occurs for $\tilde{\omega}_h/\omega<0$ or, for positive frequency modes, when,
\begin{equation} \label{SR_cond}
\omega<m\Omega_h,
\end{equation}
is satisfied.
This is the phenomenon of rotational superradiance \cite{brito2020superradiance}.

\subsubsection{WKB method} \label{sec:dbt_wkb}

Now, let us see how superradiance arises in the WKB approximation, since this will allow us to make contact with our study of vortex stability in Section~\ref{sec:vortex}.
There are two possible approaches which turn out to be equivalent in the limit of large $m$: either we apply the WKB approximation directly to \eqref{1dscat} or start from the full wave equation in \eqref{wave_eqn}.
The second approach leads to the conventional dispersion relation and as such, is the most natural to compare to the vortex analysis in Section~\ref{sec:wkb_method}.
However, we shall discuss both for completeness.

Taking the second route, we write a WKB ansatz,
\begin{equation} \label{dbt_wkb}
\phi = \mathcal{A}(r)e^{i\int p(r)dr+im\theta-i\omega t},
\end{equation}
which we then plug into the wave equation in \eqref{wave_eqn}.
Taking $\mathcal{A}'\ll p\mathcal{A}$, the leading order result is the dispersion relation $\Omega^2 = k^2$ with,
\begin{equation}
\Omega = \omega-\frac{m\ell}{r^2}+\frac{pd}{r}, \qquad k =\sqrt{p^2+\frac{m^2}{r^2}}.
\end{equation}
The dispersion relation then separates into two branches,
\begin{equation} \label{dbt_disp}
\omega_D^\pm = \frac{m\ell}{r^2} -\frac{pd}{r^2} \pm |k|,
\end{equation}
where modes with $\omega=\omega_D^+$ ($\omega=\omega_D^-$) have positive (negative) norm, as can be verified by plugging \eqref{dbt_wkb} into \eqref{norm} and discarding the dispersive terms.
At next to leading order, we obtain an expression for the amplitude $A=\sqrt{r}\mathcal{A}$,
\begin{equation}
A \propto |v_g^r\Omega|^{-\frac{1}{2}},
\end{equation}
where $v_g^r=\partial_p\omega_D^\pm$ is the radial component of the group velocity.
The dispersion relation in this case is quadratic in $p$ so there are two different solutions,
\begin{equation}
p^\pm = \frac{\tilde{\omega}d/r\pm\sqrt{-V_{|m|\gg 1}}}{1-d^2/r^2},
\end{equation}
where the term under the square root is the large $m$ limit of the potential in \eqref{dbt_pot}, which has the effect of setting $m^2-1/4+5d^2/4r^2\to m^2$.
For the remainder of this section, we drop the $|m|\gg1$ subscript and implicitly assume that this limit has been applied to $V$.
Following the procedure outlined in Section~\ref{sec:scatter}, scattering occurs when $p^\pm$ change from propagating modes into evanescent modes. This happens at the turning points $r_\tp$ which satisfy $V(r_\tp)=0$.

Let us briefly check what would happen had we applied the WKB approximation directly to \eqref{1dscat}.
Writing $R\sim A(r_*)e^{i\int P(r_*)dr_*}$ and taking $A'\ll PA$ and $P'\ll P^2$, one finds,
\begin{equation} \label{dbt_wkb2}
R\sim |V|^{-\frac{1}{4}}e^{i\int\sqrt{-V}dr_*}.
\end{equation}
Plugging this back into \eqref{decomp1} and applying $|m|\gg1$ yields precisely \eqref{dbt_wkb}, thereby confirming the equivalence of the different approaches for large $|m|$.
As a brief aside, using \eqref{dbt_wkb2} we have $P^2+V=0$. By comparing this with the $p^\pm$ modes in \eqref{pmodes} which obeyed $p^2+W_+=0$, we see that we can think of $W_+$ as the effective potential for vortex excitations.
We will revisit this point when we look at Kerr BHs later on.

\begin{figure} 
\centering
\includegraphics[width=\linewidth]{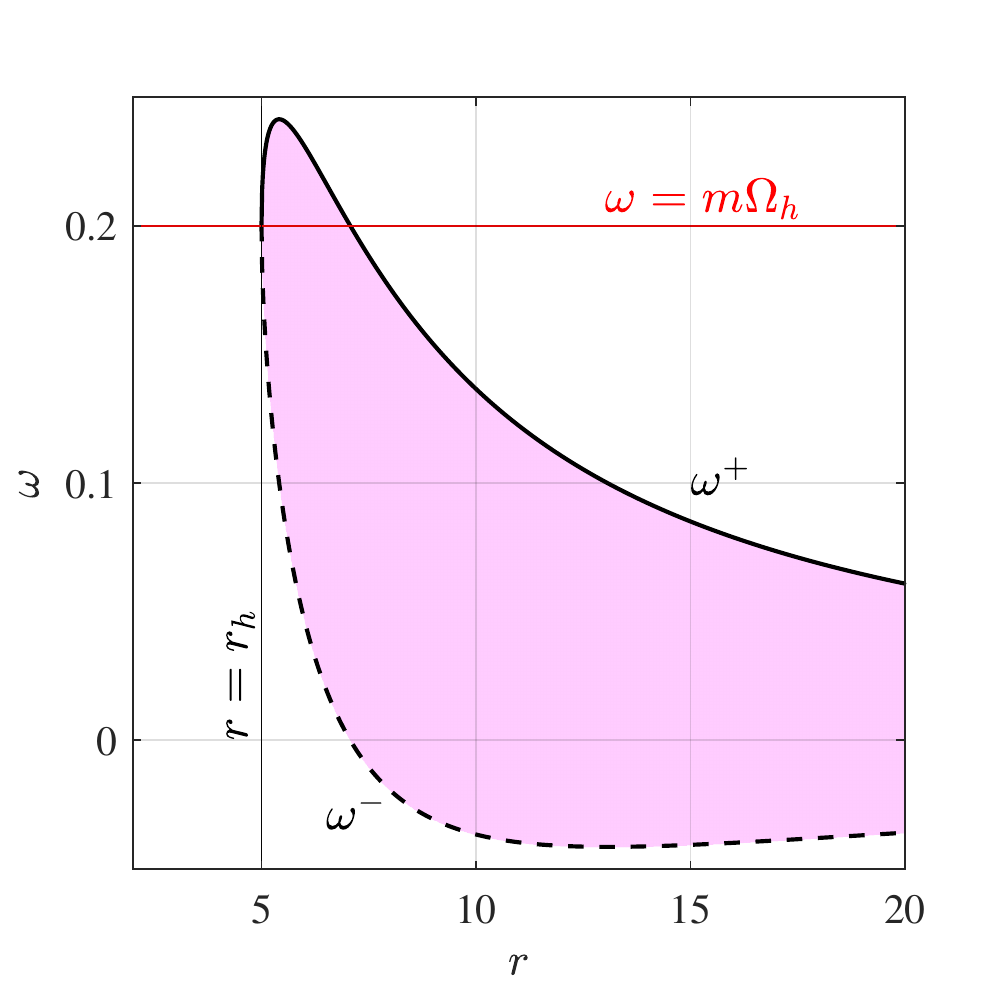}
\caption{The functions $\omega^\pm$ in \eqref{ompm_dbt} for the parameters $m=1$, $\ell=d=5$.
Superradiant amplification occurs for $0<\omega<m\Omega_h$ which is the range in which positive frequency modes tunnel to the negative norm branch of the dispersion relation.
This identifies the mechanism underlying superradiance as the same one causing the vortex instabilities.
} \label{fig:dbt_potentials}
\end{figure}

Returning to the task at hand, we now want to study scattering of the WKB modes to see how superradiance arises in this picture.
For this purpose, let us work with \eqref{dbt_wkb2}, although identical results are obtained when using \eqref{dbt_wkb} as shown in \cite{patrick2020superradiance,patrick2021rotational}.
Inspection of the form of $V$ shows that there are 0, 1 or 2 turning points for $r>r_h$.
In the case of no turning points, in-going modes are not scattered (at our level of approximation) and we have $|\mathcal{R}|=0$.
The case of 1 turning point is the marginal case where two turning points merge into one.
However, to apply the transfer matrices in \eqref{transfer} we need to have propagating modes on one side of $r_\tp$ and evanescent modes on the other.
In this case, one needs to build the transfer matrix using a different asymptotic expansion than the one we have used here, see e.g. \cite{torres2020estimate,patrick2020quasinormal}.
Proceeding to the case of 2 turning points $r_\tp\in[r_1,r_2]$, we have propagating modes for $r<r_1$ and $r>r_2$ and evanescent modes for $r_1<r<r_2$ under the potential barrier $V$.
To see where the zeros $r_{1,2}$ of the $V$ occur for different frequencies, we may write,
\begin{equation} \label{dbt_pot2}
V = -(\omega-\omega^+)(\omega-\omega^-),
\end{equation}
where these curves are $\omega^\pm = \omega_D^\pm(p_\tp)$.
Note in passing the similarity of \eqref{dbt_pot2} with the relation in \eqref{disp_potentials}, where in that case the fact that two potentials are required to make the combination on the right hand side is a consequence of dispersion.
In the current case, the $\omega^\pm$ are given by the following expressions,
\begin{equation} \label{ompm_dbt}
\omega^\pm = \frac{m\ell}{r^2} \pm \sqrt{\left(1-\frac{d^2}{r^2}\right)\frac{m^2}{r^2}}.
\end{equation}
These curves are precisely analogous to those defined in the vortex case in \eqref{potentials}.
In particular, the intersections $\omega=\cst$ and the $\omega^\pm$ give the locations of the $r_\tp$.
When $\omega$ crosses $\omega^+$ ($\omega^-$), there will be a turning point on the positive (negative) norm branch of the dispersion relation.
By looking at the example in Fig.~\ref{fig:dbt_potentials}, we see that tunnelling to the negative norm branch occurs for positive frequencies satisfying the superradiance condition in \eqref{SR_cond}.
Therefore superradiance occurs when there is tunnelling between the positive and negative norm branches of the dispersion relation.
This was exactly the mechanism that gave rise to the vortex instabilities in Section~\ref{sec:vortex_open}.
Hence, we find that superradiance and vortex instabilities are controlled by the same physics.

In conventional BH physics, relativists talk about the existence of the ergosphere as the mechanism underpinning superradiance.
It is a simple matter to see how this ties in with the WKB picture we are using, where the principal tools are the functions $\omega^\pm$.
From Fig.~\ref{fig:dbt_potentials}, it is clear that the condition $\omega^-=0$ determines the radius inside which positive frequency modes can be on the negative norm branch of the dispersion relation.
Solving this condition gives precisely the ergosphere $r_e$ defined in \eqref{re_rh}. 
Therefore the existence of an ergosphere is equivalent to the existence of a surface inside which there are $\omega>0$ modes are $\omega_D^-$, i.e. modes with negative energy density.
Note that the $m$-independence of the ergosphere happens only in the non-dispersive regime.
In the case of vortex excitations, it also makes sense to talk about an ergosphere but as an $m$-dependent notion.
Indeed, the lack of ergosphere for vortex modes above a critical $m$ is precisely the reason why those modes are stable.

As the final step, we can compute the reflection coefficient in the WKB approximation to estimate the amount of amplification caused by BH superradiance.
The method follows the scattering computation in \eqref{amp_matrix}.
However, in this case, the calculation is simpler; we only need to evaluate the scattering coefficients immediately outside the turning points since the WKB approximation guarantees that the norm of each mode is fixed as it moves off into the asymptotic regions.
We write,
\begin{equation} \label{amp_matrix2}
\begin{pmatrix}
A^+_1 \\ A^-_1
\end{pmatrix} = \left|\frac{V(r_2)}{V(r_1)}\right|^\frac{1}{4}
T\begin{pmatrix}
0 & e^{S_{12}} \\ e^{-S_{12}} & 0
\end{pmatrix}\widetilde{T}\begin{pmatrix}
A^+_2 \\ A^-_2
\end{pmatrix},
\end{equation}
with,
\begin{equation}
S_{ij} = \int^{{r_*}_j}_{{r_*}_i}|V|^\frac{1}{2}dr_*.
\end{equation}
Assuming $\omega>0$, the absorbing BC at the horizon fixes $A_2^+=0$ for $\tilde{\omega}_h>0$ and $A_2^-=0$ for $\tilde{\omega}_h<0$.
To solve for the reflection coefficient, we then note that $|V(r_2)|^\frac{1}{2}A_2^j = |V(\infty)|^\frac{1}{2}A_\infty^j$ and we apply the definition in \eqref{scat_coefs} to obtain,
\begin{equation} \label{refl_wkb}
\mathcal{R} = e^{-\frac{i\pi}{2}}\left(\frac{4-e^{-2S_{12}}}{4+e^{-2S_{12}}}\right)^{\frac{\tilde{\omega}_h}{\omega}}.
\end{equation}
As expected, this is less than 1 for $\tilde{\omega}_h>0$ and greater than 1 for $\tilde{\omega}_h<0$.

In summary, our analysis demonstrates that superradiance occurs when in-coming positive norm modes tunnel to the negative norm branch of the dispersion relation.
The comparison between Figs.~\ref{fig:potentials} and \ref{fig:dbt_potentials} makes the connection to the stability of quantised vortices apparent.
With this connection, the vortex instability can be interpreted as a superradiating bound state trapped in the vortex core.
Alternatively, with our understanding of vortex instabilities in closed systems, superradiance can be interpreted as a coupling between positive norm modes in the asymptotic region and negative norm modes near the horizon, which act to lower the energy of the BH. 
Unlike the vortex where this coupling only occurred for special system sizes, the spectrum of BH modes is continuous due to the open boundary conditions and hence, the coupling always occurs in the frequency band given by \eqref{SR_cond}.
Finally, although we have neglected dispersion, note that the same picture emerges when taking dispersion into account, as has recently been demonstrated in \cite{patrick2021rotational}, although the superradiance criterion in \eqref{SR_cond} picks up corrections in that case.

\subsubsection{Stability} \label{sec:drain_stab}

In the (analogue) BH case, incident modes can be amplified but no instability occurs.
This is due to the presence of absorbing BCs at the horizon and infinity which means that perpetual amplification is prevented.
In the vortex case, there was a reflective BC at the origin which allowed for the amplified mode to become trapped.
This suggests that adding a generic drain component to a vortex will stabilise it, since the drain can act like a horizon and dissipate away the negative norm mode in the core.
One might then reasonably ask what happens if the drain is not perfectly absorbing and instead a small amount of reflection occurs.
In this section, we search for the conditions under which a quantised vortex can be stabilised by a drain.
This is also of interest in view of recent reports that multiply-quantised vortices are stable in polariton-fluid experiments when a drain is added to the flow \cite{alperin2021multiply}.

Let us assume that we have an open system extending to infinity and that the location $r_h$ is no longer a true horizon in the sense that modes are reflected there with reflection coefficient,
\begin{equation}
\mathcal{Z} = A_h^-/A_h^+.
\end{equation}
Here we have restricted ourselves to looking at modes with $\tilde{\omega}_h<0$ (i.e. superradiant ones) so that $A_h^+$ is the in-going mode and $A_h^-$ is the out-going one.
We can now perform our scattering computation, taking the matrix equation in \eqref{amp_matrix2} as our starting point.
Then applying $A^-_2=0$ (since we send no waves into the system) and transporting the modes at $r_1$ to the horizon through $A_h^\pm = |V(r_1)/V(r_h)|^\frac{1}{4}e^{\mp iS_{h1}}A^\pm_1$, we obtain the following resonance condition,
\begin{equation}
e^{2iS_{h1}} + \mathcal{R}\mathcal{Z} = 0,
\end{equation}
where the $\mathcal{R}$ is the superradiant expression in \eqref{refl_wkb}.
If the growth/decay rate is small compared to the oscillation frequency, i.e. $|\Gamma|\ll|\omega|$, this splits into,
\begin{equation} \label{res_dbt}
S_{h1} = \pi\left(\nu+\tfrac{1}{4}\right) + \chi, \qquad \Gamma = \frac{\log\left|\mathcal{R}\mathcal{Z}\right|}{2\left|\partial_\omega S_{h1}\right|},
\end{equation}
where $\nu=0,1,...$ indexes the different possible bound states and we have written $\mathcal{Z}=|\mathcal{Z}|e^{2i\chi}$.
Note that the formulae in \eqref{res_open2} are recovered for the case with $\mathcal{Z}=i$.

From the expressions in \eqref{res_dbt} we learn that for the system to be stable (i.e. $\Gamma<0$), the product of the reflection coefficients needs to be less than unity, i.e. 
\begin{equation} \label{stable_cond}
|\mathcal{R}\mathcal{Z}|<1.
\end{equation}
Recalling we are in the superradiant case where $|\mathcal{R}|>1$, this gives us the intuitive result that the system is stable if the boundary at $r_h$ absorbs the wave more than the barrier at $r_1$ amplifies it.
Repeating the same analysis for the non-superradiant case, one finds the same stability condition as in \eqref{stable_cond}, which is always satisfied if the boundary at $r_h$ is an absorber and not an amplifier.

There are a couple of reasons why one can expect the condition in \eqref{stable_cond} to be satisfied in a realistic experiment.
Firstly, the reflection coefficient for amplified modes in BH-type rotational superradiance is not usually vastly greater than unity, see e.g. \cite{richartz2015rotating,demirkaya2020acoustic,patrick2020superradiance,patrick2021rotational,brito2020superradiance}.
By contrast, an efficient drain would be expected to highly damp out in-coming modes, e.g. in \cite{torres2020estimate} the maximum value of $\mathcal{Z}$ was $\sim 0.2$ for counter-rotating modes.
Lastly, dissipation inherent in any real experiment (due to e.g. viscosity or conversion into other types of waves \cite{torres2017rotational}) will further suppress the amount of superradiant amplification, such that it becomes more difficult for $|\mathcal{R}\mathcal{Z}|$ to exceed unity.

As a final note, we point out that here we have only considered the possibility of unstable modes in the case of an infinite system, i.e. ergoregion instabilities.
If the system were finite, it would also be possible to have instabilities outside the ergoregion, i.e. black hole bomb modes.
In the next section, we discuss this type of instability in the context of an astrophysical black hole.

\subsection{Kerr spacetime} \label{sec:kerr}

Finally, we come to study superradiance around an astrophysical Kerr BH.
In this section, we will focus on a superradiant instability that occurs for massive fields in the Kerr spacetime, comparing and contrasting this to the instability around quantised vortices.
We work in natural units such that $G=\hbar=c=1$, where $G$ is the gravitational constant, $\hbar$ is the reduced Planck's constant and $c$ is the speed of light.

\subsubsection{The metric}

The Kerr metric describes a rotating BH spacetime with mass parameter $M$ and angular momentum $J=aM$.
The popular representation of the metric involves Boyer-Lindquist coordinates $x^\nu=(t,r,\varphi,\theta)$ where $\nu$ is a spacetime index and, in contrast to previous sections where only two spatial dimensions were considered, $\theta\in[0,\pi]$ is the polar angle and $\varphi\in[0,2\pi)$ is the azimuthal angle.
Note in particular that in this section $\varphi$ is a coordinate and not a field.
The Kerr metric $g_{\nu\sigma}$ in these coordinates is \cite{visser2007kerr},
\begin{equation} \label{Kerr}
g_{\nu\sigma} dx^\nu dx^\sigma = -\frac{\Delta_r}{\rho^2}\hat{\omega}_1^2 + \frac{a^2\sin^2\theta}{\rho^2}\hat{\omega}_2^2 + \frac{\rho^2}{\Delta_r}dr^2 + \rho^2d\theta^2,
\end{equation}
with the metric components,
\begin{equation}
\hat{\omega}_1 = dt-a\sin^2\theta d\varphi, \qquad \hat{\omega}_2 = dt - \frac{r^2+a^2}{a}d\varphi,
\end{equation}
The metric functions are,
\begin{equation}
\Delta_r = r^2+a^2-2Mr, \qquad \rho^2 = r^2 + a^2\cos^2\theta,
\end{equation}
where $\rho$ should not be confused with the density in earlier sections.
The inner and outer horizons, $r_-$ and $r_+$ respectively are the locations where $\Delta_r=0$,
\begin{equation}
r_\pm = M \pm\sqrt{M^2-a^2}.
\end{equation}
Since we are only concerned with the outer horizon from here on, let $r_h=r_+$ for labelling purposes.
To avoid having a naked singularity, the rotation parameter must satisfy $a<M$.
The ergosphere is the region inside which an observer is forced to co-rotate with the BH with respect to an observer located at infinity \cite{visser2007kerr}.
It is located inside $r_h<r<r_e$ where,
\begin{equation}
r_e = M+\sqrt{M^2-a^2\cos^2\theta}.
\end{equation}
From here on, we will work in units where $M=1$ so that $a<1$.

\subsubsection{The Klein-Gordon field}

The equation of motion for a massive scalar field $\phi$ with mass $\mu$ is the Klein-Gordon (KG) equation,
\begin{equation} \label{KGeqn}
\frac{1}{\sqrt{-g}}\partial_\nu\left(\sqrt{-g}g^{\nu\sigma}\partial_\sigma\phi\right) - \mu^2\phi = 0,
\end{equation}
where $g^{\nu\sigma}$ and $g$ are the inverse and determinant of the metric respectively, and repeated indices are summed over in accordance with the Einstein summation convention.
Since $t$ and $\varphi$ are symmetries of the metric (or in the language of GR, $\partial_t$ and $\partial_\varphi$ are Killing vectors of the spacetime) the field $\phi$ is split into the different $\omega$ and azimuthal $m$ components,
\begin{equation} \label{kerr_decomp}
\phi = \int d\omega \sum_m R_{\omega m}(r) S_{\omega m}(\theta)e^{im\varphi-i\omega t},
\end{equation}
and KG equation splits into two separate equations.
The angular part of the field is governed by,
\begin{equation} \label{S_eq}
\begin{split}
& \sin\theta\partial_\theta\left(\sin\theta\partial_\theta S_{\omega m}\right) -\big[\left(\omega a \sin^2\theta -m\right)^2 \\ 
& \qquad \quad + \sin^2\theta\left(\mu^2a^2\cos^2\theta - j_{\omega m}\right)\big] S_{\omega m} = 0,
\end{split}
\end{equation}
and the radial part obeys,
\begin{equation} \label{Kerr_R_eq}
\Delta_r\partial_r\left(\Delta_r\partial_r R_{\omega m}\right) + K^2 R_{\omega m} = 0,
\end{equation}
with,
\begin{equation} \label{K_pot}
K^2 = \left[\omega(r^2+a^2)-am\right]^2 - \Delta_r\left(j_{\omega m}+\mu^2 r^2\right).
\end{equation}
The separation constant $j_{\omega m}$ can be obtained as the eigenvalue of \eqref{S_eq} when imposing regular BCs at the poles $\theta=\pm\pi/2$.
Indeed, as well as solving numerically, one can apply the by now familiar WKB methods to \eqref{S_eq} to find approximate expressions for the $j_{\omega m}$ that give some insight into it's behaviour.
Such an analysis can be found in Appendix~\ref{app:sep}.
In the non-rotating $a=0$ case, we have $j_{\omega m}=l(l+1)$, where $l\geq|m|$ is a positive integer called the polar ``quantum number''.
In the rotating case, $l$ can also be used to index the different $j_{\omega m}$.
From here on, we assume that \eqref{S_eq} has been solved and the value of $j_{\omega m}$ determined.

Starting from \eqref{Kerr_R_eq}, one can follow a similar analysis to that outlined in Section~\ref{sec:SR_con} to deduce the occurrence of superradiant amplification of incident $\omega>0$ modes in the frequency band,
\begin{equation} \label{Kerr_SR}
\mu<\omega<m\Omega_h, \qquad \Omega_h = \frac{a}{r_h^2+a^2},
\end{equation}
where $\Omega_h$ is the angular velocity of the spacetime on the horizon \cite{visser2007kerr}.
Since we have already drawn the comparison between superradiance and the vortex instability mechanism using the analogue BH, we will not reproduce the full analysis here (the interested reader can find a full treatment in e.g. \cite{unruh1974second,ford1975quantization}). 
Instead, here we choose to focus on another aspect of the Kerr spacetime particular to massive fields: the BH bomb instability.

\subsection{Bound states of a Kerr BH} \label{sec:kerr_BS}

The possibility of having (quasi-)bound states around the BH arises due to the non-zero rest mass $\mu\neq0$ of the field, which acts as a mirror at large $r$ reflecting the out-going part of the field back into centre.
If this happens to a superradiant mode, then the amplified mode is reflected back into the BH where it is further amplified, leading to an instability.
In the literature, an instability of this type is known as a BH bomb \cite{brito2020superradiance}.
We shall shortly see that although, like the vortex instability, these instabilities owe their existence to the coupling of positive norm modes to negative norm ones, they are unlike the vortex one in the sense that they occupy the positive norm region of the system.
These unstable BH frequencies have been computed using precise numerical algorithms in e.g. \cite{dolan2007instability}.
For the purposes of this work, a WKB approximation of the modes will be sufficient since this permits an easy comparison with the vortex case.

\subsubsection{WKB method}

To apply the WKB method to \eqref{Kerr_R_eq}, we write,
\begin{equation} \label{Kerr_wkb}
R_{\omega m}(r) = \mathcal{A}(r)\,e^{\pm i\int K(r)\,\Delta_r^{-1}dr},
\end{equation}
for $\Delta_r\partial_rK\ll K^2$ and $\Delta_r\partial_r\mathcal{A}\ll K\mathcal{A}$.
The leading order equation is then exactly \eqref{K_pot} and at next to leading order we find $\mathcal{A}\sim|K|^{-\frac{1}{2}}$.
It is useful at this point to introduce the effective radial potential,
\begin{equation} \label{Kerr_pot}
\begin{split}
V = & \ -K^2/(r^2+a^2)^2, \\
 = & \ -\left(\omega-\frac{ma}{r^2+a^2}\right)^2 + \frac{\Delta_r}{(r^2+a^2)^2} \\
 & \qquad \qquad \qquad \qquad \qquad \times \left(j_{\omega m} + \mu^2r^2\right),
\end{split}
\end{equation}
which is analogous to the large $|m|$ limit of the draining vortex potential \eqref{dbt_pot} in the sense that \eqref{Kerr_pot} is the WKB limit (in this case large $l$) of the radial scattering potential arising when not working in the WKB approximation.
The turning points $r_\tp$ are the locations where $V(r_\tp)=0$; for $V<0$ the modes are propagating whereas for $V>0$ they will be evanescent.
In particular, for $\omega^2<\mu^2$ the solution is evanescent at infinity.

\subsubsection{Boundary conditions}

The asymptotics of the solutions can be found by converting to the tortoise coordinate $r_*$ defined by,
\begin{equation} \label{tortoise2}
dr_* = \frac{r^2+a^2}{\Delta_r}dr.
\end{equation}
Using this coordinate, the WKB solution in \eqref{Kerr_wkb} may be written,
\begin{equation} \label{Kerr_wkb2}
R_{\omega m}\sim (r^2+a^2)^{-\frac{1}{2}}|V|^{-\frac{1}{4}}e^{\pm i\int\sqrt{-V}dr_*},
\end{equation}
which is notably similar to \eqref{dbt_wkb2}. 
The extra factor at the front is because in this case we have $A=\sqrt{r^2+a^2}\mathcal{A}$ (rather than the relation we had for the 2D vortex which was $A=\sqrt{r}\mathcal{A}$).

Plugging in the limiting form for $V$ in the two limits $r\to r_h$ and $r\to\infty$, modes with $\omega^2<\mu^2$ become,
\begin{equation} \label{R_asymp}
R_{\omega m} \sim \frac{1}{\sqrt{r^2+a^2}}\begin{cases}
A_h e^{-i\tilde{\omega}_h r_*}, \ \quad \qquad r\to r_h \\
A^\downarrow_\infty e^{-\sqrt{\mu^2-\omega^2}r_*}, \quad r\to\infty
\end{cases}
\end{equation}
where we have adopted the labelling convention established in Section~\ref{sec:scatter}.
We have also defined,
\begin{equation} \label{hor_freq_kerr}
\tilde{\omega}_h = \omega-m\Omega_h,
\end{equation}
for $\Omega_h$ in \eqref{Kerr_SR}.
The BC at spatial infinity has been imposed by discarding the non-physical growing mode and on the horizon we have imposed the absorbing BC by allowing only the in-going mode, which in terms of the WKB solution in \eqref{Kerr_wkb2} is the one with the $-$ in the phase for $\tilde{\omega}_h>0$ and $+$ in the phase for $\tilde{\omega}_h<0$.
Therefore the BC for the (quasi-)bound states of the Kerr spacetime can be summarised as,
\begin{equation} \label{BS_BCs}
A^\epsilon_h = A^\uparrow_\infty = 0
\end{equation}
with $\epsilon=\sgn(\tilde{\omega}_h)$.

\subsubsection{The resonance condition}

The BCs for \eqref{R_asymp} are only satisfied only for a discrete set of frequencies.
To obtain a condition for these frequencies, consider the following argument.

Since the solution propagates on the horizon and decays toward spatial infinity there must be at least one turning point $r_\tp$ in the intermediate region to convert propagating modes into evanescent modes. 
However, if there is only a single turning point, then the in-going mode at $r_h$ will be completely reflected as it is traced back to $r_\tp$ and there will also be an out-going mode on the horizon, which contradicts the BC.
The next possibility is that there are an extra two turning points located somewhere between the horizon and first turning point.
Let these three be labelled $r_1<r_2<r_3$ such that the solution propagates for $r<r_1$ and $r_2<r<r_3$, and is evanescent for $r_1<r<r_2$ and $r>r_3$.
The existence of three (real) turning points is a necessary (but not sufficient) condition for a solution which satisfies \eqref{BS_BCs}.
Note that the next possibility up, i.e. 5 turning points, is not possible since $K^2$ is a quartic polynomial and has only 4 roots (one of which is at negative $r$ and therefore not in the coordinate range).

To find the resonance condition, we perform another scattering computation,
\begin{widetext}
\begin{equation} \label{amp_matrix3}
\begin{pmatrix}
A^+_1 \\ A^-_1
\end{pmatrix} = \left|\frac{V(r_3)}{V(r_1)}\right|^\frac{1}{4}
T\begin{pmatrix}
0 & e^{S_{12}} \\ e^{-S_{12}} & 0
\end{pmatrix}\widetilde{T}\begin{pmatrix}
e^{-iS_{23}} & 0 \\ 0 & e^{S_{23}}
\end{pmatrix}T\begin{pmatrix}
A^\downarrow_3 \\ A^\uparrow_3
\end{pmatrix},
\end{equation}
\end{widetext}
with,
\begin{equation} \label{S_kerr}
S_{ij} = \int^{r_j}_{r_i}|K|\,\frac{dr}{\Delta_r}.
\end{equation}
Since we are interested in the unstable modes (which can only exist when there is superradiance) we take the superradiant BCs in \eqref{BS_BCs} to get $A^-_1=A^\uparrow_3=0$.
Inserting these into \eqref{amp_matrix3} we solve to find the resonance condition,
\begin{equation} \label{res_cond}
e^{2iS_{23}}+|\mathcal{R}|^{-1}=0,
\end{equation}
where the reflection coefficient $\mathcal{R}$ is the same as the one defined in \eqref{refl_wkb} but using the definition of $\tilde{\omega}_h$ and $S_{ij}$ in \eqref{hor_freq_kerr} and \eqref{S_kerr} respectively. 
This condition is solved by complex frequencies $\omega_\mathbb{C}=\omega+i\Gamma$.
Assuming $|\Gamma|\ll|\omega|$ splits \eqref{res_cond} in two in the usual fashion,
\begin{equation} \label{BohrSomm}
S_{23}(\omega_\nu) = \pi\left(\nu+\tfrac{1}{2}\right), \qquad \Gamma_\nu = \frac{\log|\mathcal{R}(\omega_\nu)|}{2\partial_\omega S_{23}(\omega_\nu)},
\end{equation}
where $\nu=0,1...$ indexes the different bound states.
The expressions in \eqref{BohrSomm} describe modes which are in a cavity for $r\in[r_2,r_3]$ located outside the potential barrier at $r\in[r_1,r_2]$. 
As verified in Appendix~\ref{app:norm}, these modes have positive norm inside the cavity and negative norm as they transmit into the horizon.
They also no longer radiate to infinity as in the vortex case because the field mass acts as a perfect mirror and reflects them back in.
Since modes in \eqref{BohrSomm} are in a superradiant frequency band, each time they scatter at $r_2$ they will be superradiantly amplified, leading to perpetual amplification.
The negative norm mode which is transmitted through the barrier then dissipates into the horizon, lowering the mass and angular momentum of the BH as with normal superradiant scattering.

From \eqref{BohrSomm}, we can already draw comparisons with the vortex case.
Comparing with \eqref{res_open2}, and realising in that case $X=|\mathcal{R}|$, the most apparent difference is the minus sign in front of the expression for $\Gamma$, which is absent for the present case.
This is because the term in the denominator of $\Gamma$ has the same sign as the norm of the mode in the cavity, which is positive in the Kerr case and negative for the vortex.
Hence, the overall effect is that the sign of $\Gamma$ coincides with that of $\log|\mathcal{R}|$.

\begin{figure} 
\centering
\includegraphics[width=\linewidth]{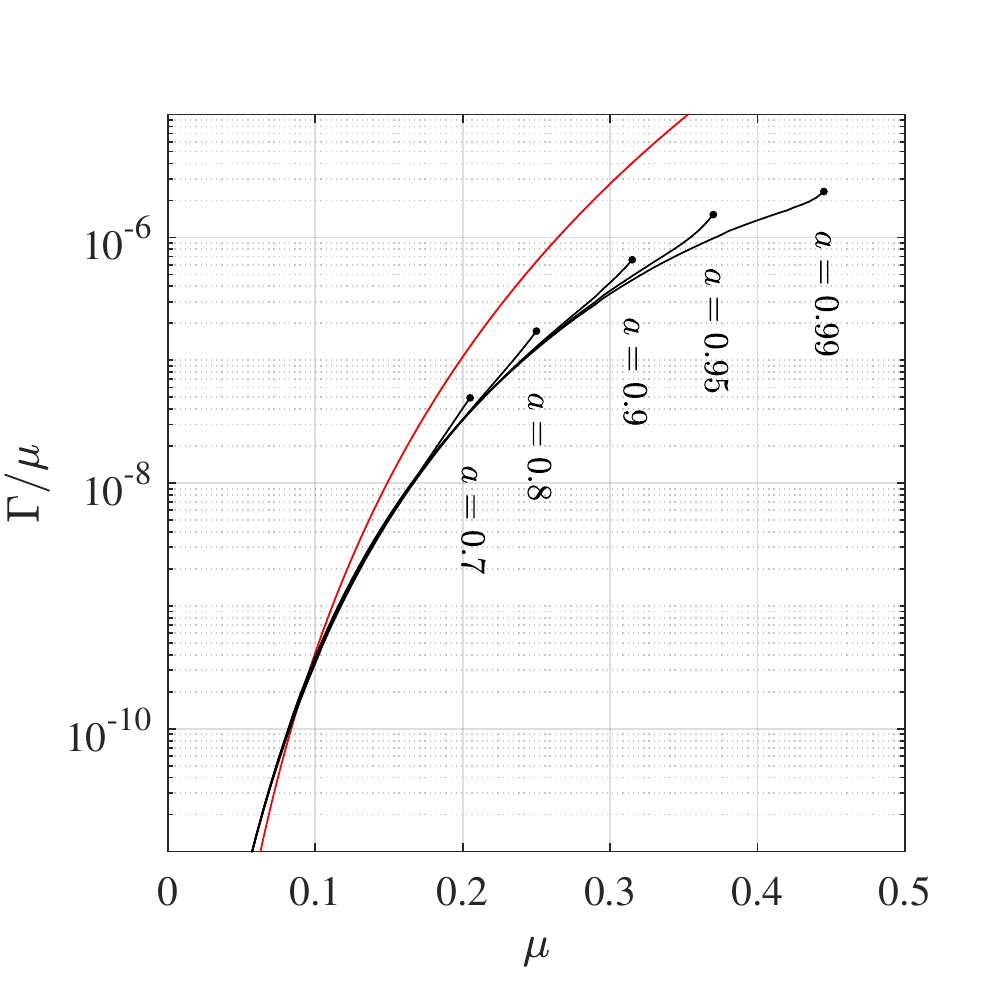}
\caption{Growth rates (black lines) for the $m=l=1$ mode.
The black dots indicate the value of $\mu$ above which the instability disappears.
The red line is the small $\mu$ approximation $\Gamma\sim\mu^9/24$ found in \cite{zouros1979instabilities}.
} \label{fig:rates1}
\end{figure}

The instability rates predicted by \eqref{BohrSomm} are plotted as a function of the mass coupling for the $m=l=1$ mode for several values of $a$ in Fig.~\ref{fig:rates1}.
In particular, the most unstable mode (which is the one with the largest $\Gamma$) occurs at a mass coupling just below $\mu\sim 0.5$ when the BH is almost maximally rotating.
These findings agree qualitatively with precise numerical computations in the literature, see e.g. Fig.~6 of \cite{dolan2007instability}, with the approximation used here having a tendency to over-estimate the size of $\Gamma$.
This is expected since the WKB method is known to overestimate the reflection coefficient in similar calculations, e.g. \cite{patrick2021rotational}.

\subsubsection{Hydrogenic spectrum} \label{sec:hydrogen}

It is well known that the frequency spectrum of the Kerr bound states has a similar form to the electronic spectrum of the Hydrogen atom.
Here, we will show how this analogy can be derived from \eqref{BohrSomm}.

The derivation begins by noting that as $\omega\to\mu$ from below, the turning point $r_3$ is sent to infinity,
\begin{equation} \label{r3_div}
r_3\overset{\omega\to\mu}{\sim}\frac{2\mu^2}{\mu^2-\omega^2}.
\end{equation}
Since the cavity becomes infinitely wide in this limit, we would expect to be able to fit an infinite number of states inside of it.
For $\omega\sim\mu$, the function $K^2$ is well approximated by discarding the term which is linear in $r$ and the constant term, and the zeros $r_{2,3}$ can be approximated as the two zeros of the truncated $K^2$.
Using $\Delta_r\sim r^2$ for large $r$, the integral $S_{23}$ in \eqref{res_cond} is,
\begin{equation} \label{hydro1}
S_{23}(\omega)\approx \frac{\pi\mu^2}{\sqrt{\mu^2-\omega^2}} - \pi\left(l+\tfrac{1}{2}\right),
\end{equation}
where a detailed derivation of this result can be found in Appendix~\ref{app:sep}.
Combining \eqref{hydro1} with the condition \eqref{BohrSomm}, the spectrum of bound state frequencies close to $\mu$ is approximately given by,
\begin{equation} \label{hydro_sec}
\omega_\nu = \mu\left(1-\frac{\mu^2}{2\bar{\nu}^2}\right),
\end{equation}
where the ``principle quantum number'' $\bar{\nu}=\nu+l+1$ is assumed large, since $\nu$ is large as $\omega\to\mu$ and the WKB approximation assumes large $l$.
This has precisely the form of the (semi-classical) electronic spectrum of the Hydrogen atom \cite{detweiler1980klein}.
Note that \eqref{hydro_sec} can also be found by approximating the KG equation at large $r$, which is then exactly in the form of the Schr\"odinger equation for a charged particle moving in a Coulomb potential, e.g. \cite{fox2006quantum,detweiler1980klein,dolan2007instability}.

\subsubsection{Comparing the Kerr and vortex potentials} \label{sec:comp}

\begin{figure*} 
\centering
\includegraphics[width=\linewidth]{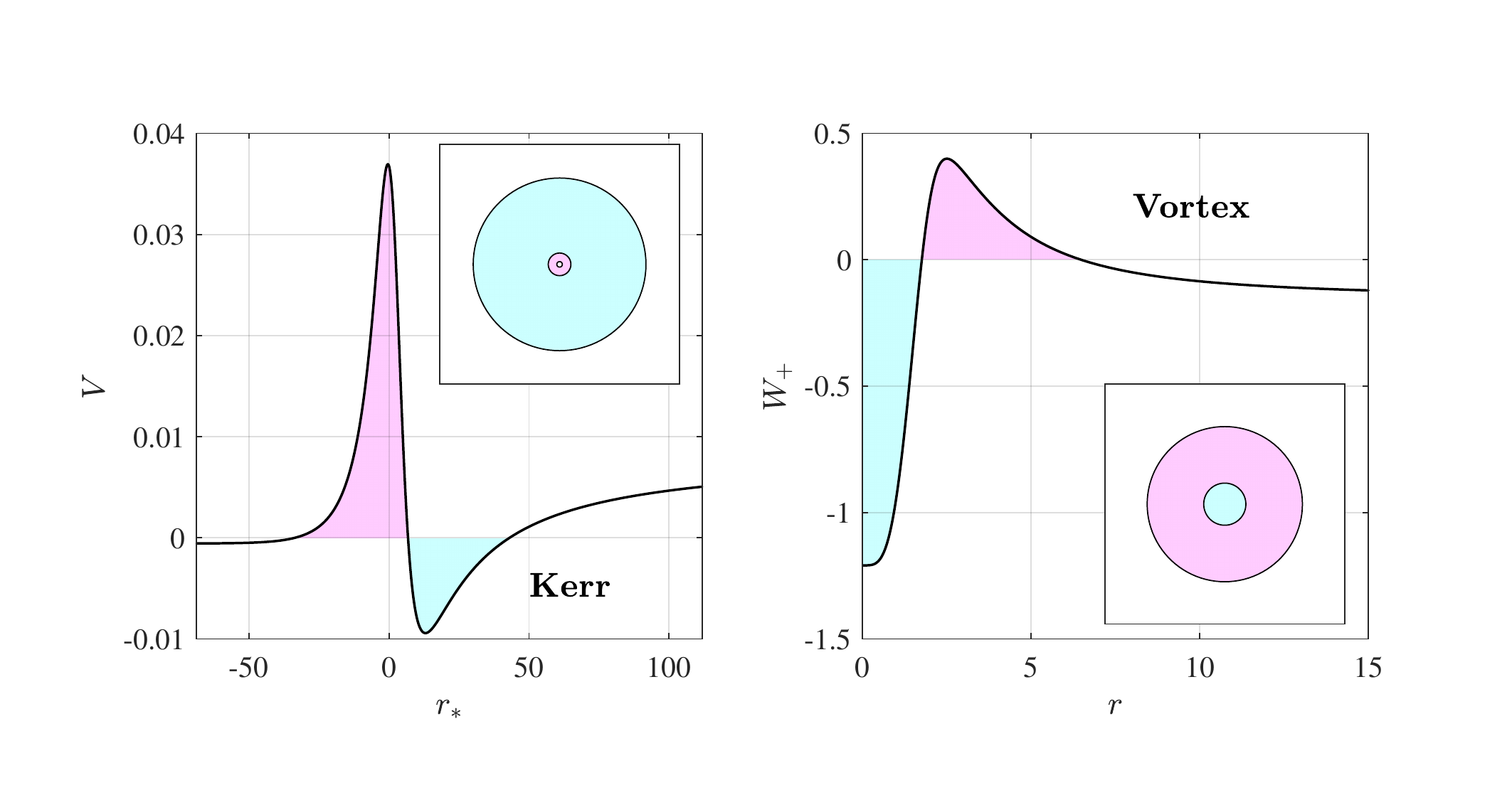}
\caption{Comparison of the radial potentials for the Kerr spacetime and the quantised vortex.
For the Kerr diagram, the parameters are $\omega=0.4075$, $m=1$ and $\mu=0.42$ and $a=0.99$, whereas for the vortex, they are $\omega=0.3954$, $m=2$ and $\ell=2$.
These parameters correspond to an instability in both cases.
In these potentials, the modes have negative norm to the left of the peaks of $V$ and $W_+$ and positive norm to the right.
The instability occupies the cavity (blue region).
In both cases, the oscillation frequency is determined by the phase integral over the blue region, that is, \eqref{res_open2} for the vortex and \eqref{BohrSomm} for the BH.
The growth rate is given by barrier integrals $S_{12}$ appearing in those equations, which are the related to the pink areas.
The difference between the two systems is that the Kerr instability lives in the positive norm region of the flow and tunnels into the BH whereas the vortex instability lives in the negative norm region and tunnels out to infinity.
The insets indicate the corresponding 2D profiles on a Cartesian grid.
} \label{fig:comp}
\end{figure*}

Already from the resonance conditions in \eqref{res_open2} and \eqref{BohrSomm}, we can see that the features of the different instabilities are determined in a similar manner by the properties of the two systems.
To expand upon the comparison, we compare the radial potential for the BH and the vortex  to see explicitly where in the system the instability becomes trapped.

In the BH case, the potential is the one in \eqref{Kerr_pot} which, as seen from \eqref{Kerr_wkb2}, obeys $p_*^2+V=0$ where $p_*$ is the component of the wavevector in the tortoise coordinate $r_*$.
In the vortex case, the curves $\omega^\pm$ appear as the more natural objects to determine the locations where scattering occurs in the system.
However, as seen from \eqref{pmodes}, and also discussed around \eqref{disp_potentials}, the function $W_+$ obeys $p^2+W_+=0$ and can therefore be thought of as an effective potential.
$W_+$ is therefore a suitable object to compare with the Kerr potential $V$.

In Fig.~\ref{fig:comp}, the two potentials are displayed for particular choices of parameters corresponding to an instability.
These plots paint a similar picture of the BH and non-draining vortex instabilities. 
Namely, the unstable mode occupies a cavity in the effective potential (blue region) where it grows over time, then tunnels through the barrier (pink region).
In both cases, the area in the blue determines the oscillation frequency and the pink area gives the growth rate (really it is area of the same region under the square root of the black curve which gives the barrier integrals $S_{12}$).
The plots also help understand the key differences.
In both cases, the modes have positive norm to the right of the barrier and a negative norm to the left.
Therefore, the Kerr instability lives in the positive norm region of the spacetime outside the ergosphere and tunnels into the BH.
By contrast, the vortex instability occupies the centre of the system where it has a negative norm.

Finally, this picture provides a natural interpretation of the different dynamics resulting from the two instabilities.
The Kerr instability saps rotational energy from the system, spinning the BH down until it is no longer rotating, leaving the scalar field to eventually dissipate through the emission of gravitational waves \cite{east2018massive}.
This is made possible since the instability lives outside the BH.
By contrast, the vortex instability grows in the centre of the system, forcing the $\ell$-charged vortex to fragment into $\ell$ unit charged vortices which are pushed apart, e.g. \cite{shin2004dynamical,isoshima2007spontaneous,okano2007splitting} (see also our companion paper where we study this splitting for $\ell=2$ \cite{patrick2021origin}).
In this sense, the vortex is torn apart from the inside.

\subsubsection{Analogy with $\alpha$-decay} \label{sec:alpha}

In Section~\ref{sec:hydrogen}, we saw that the spectrum of BH bound states in the limit $\omega\to\mu$ have an analogue in atomic physics, namely, the electronic spectrum of the Hydrogen atom.
It is natural then to enquire as to whether the vortex instabilities also have an atomic analogue.
Having seen in Fig.~\ref{fig:comp} that Kerr instabilities live outside the BH whereas vortex instabilities live inside the core, one might expect that the vortex states should bear some resemblance to certain nuclear excitations.
We show that this expectation is in a sense correct, using the example of $\alpha$-decay.

The phenomenon of $\alpha$-decay involves a radioactive isotope that fragments into an $\alpha$-particle (i.e. two protons + two neutrons with charge $2e$, where $e$ is elementary charge,  and mass $M_\alpha$) and a daughter nucleus (charge $Z_ne$, where $Z_n$ is the proton number, and mass $M_n$).
It is known that the lifetime of the radioactive species is inversely proportional to the square root of the energy of the ejected particle, which is the so-called Geiger-Nutall law \cite{geiger1911lvii,ren2012new}.
The following simple model was proposed by Gamow \cite{gamow1928quantentheorie} to explain this dependence.
The Schr\"odinger equation for the two particle system can be reduced to,
\begin{equation}
i\hbar\partial_t\psi=-\frac{\hbar^2}{2\mu}\nabla^2\psi + V(r)\psi,
\end{equation}
where $\mu=(M_\alpha M_n)/(M_\alpha+ M_n)$ is the reduced mass, $V(r)$ is a central potential and $r$ is the distance from the centre of mass.
Gamow's model is based on a simple potential in which the Coulomb interaction acts to repel the particles at large $r$ whereas the nuclear force leads to a strong attraction at small $r$.
In particular,
\begin{equation} \label{gamow_pot}
V = -V_0\, \Theta(r_1-r) + \frac{q}{r}\, \Theta(r-r_1),
\end{equation}
where $\Theta(r)$ is the step function, $q=2Z_n e^2>0$ is the charge coupling and $r_1$ is the radius inside which the nuclear attraction dominates.
The attractive nuclear force is assumed to have a fixed value of $V_0=\cst$.
This has a potential well in the region $0<r<r_1$ inside which particles with $E<V(r_1)$ can become trapped.
If the particle has $E>0$, it can escape to infinity with a probability determined by the width of the barrier.
Decomposing the wavefunction as $\psi=r^{-1}R(r)Y^m_l(\theta,\varphi)e^{-iEt/\hbar}$ with $Y^m_l$ a spherical harmonic, this becomes,
\begin{equation}
\partial_r^2 R(r) = \frac{l(l+1)}{r^2}R + (V-E)R,
\end{equation}
where we have scaled by the reference length scale $r_0=\hbar/\sqrt{2V_0\mu}$, which amounts to setting $\hbar,\mu$ and $V_0$ equal to 1.
In the following, we consider only the $l=0$ mode for simplicity.

\begin{figure} 
\centering
\includegraphics[width=\linewidth]{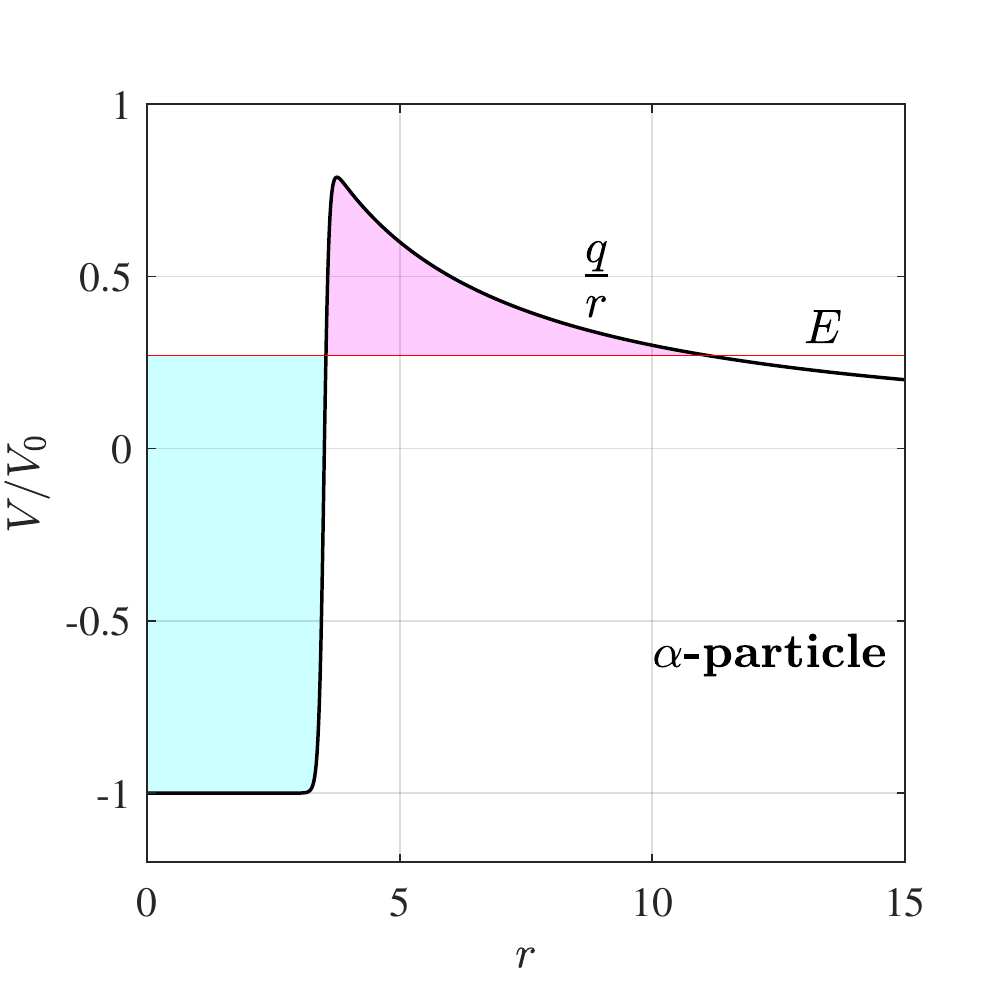}
\caption{Example of the potential in \eqref{gamow_pot} for $V_0=1$, $q=3$, $r_1=3.5$ and $\Theta(r)\to\tanh(10\,r)$.
There is a clear similarity with the vortex potential shown on the right panel of Fig.~\ref{fig:comp}.
In this sense, the confinement of the vortex instability is analogous to the nuclear resonance in $\alpha$-decay.
} \label{fig:gamow}
\end{figure}

The excited states in the nucleus and their associated lifetimes can be estimated using the WKB approximation.
In order to employ our method, we require that $V$ be a smooth function at the turning points.
We therefore make the replacement $\Theta(r)\to\tanh(\kappa r)$ in \eqref{gamow_pot}, where $\kappa$ is a large number controlling the steepness of the step.
We show an example for $V$ in Fig.~\ref{fig:gamow}, taking $\kappa=10$.
Writing $R\sim |p|^{-\frac{1}{2}}\exp(i\int p\,dr)$ with $p^2+V=E$, the WKB matrix calculation is equivalent to the one shown in \eqref{amp_matrix}, with $Q=p$ and $S_{ij} = \int^{r_j}_{r_i}|p|dr$.
The four turning points $r_i$ are $r_0=0$, $r_2\simeq q/E$ (provided $\kappa$ is large enough), $r_\infty=\infty$ and $r_1$ is a free parameter.
To obtain the resonance condition, we apply the no-radiation BC ($A^-_\infty=0$) and take a Neumann BC at the origin ($A^-_0=A^+_0$).
In the limit $e^{-S_{12}}\ll 1$, the complex eigenvalue $E+i\Gamma$ is given by,
\begin{equation} \label{alpha_res}
\cos\left(S_{01}(E)+\frac{\pi}{4}\right) = 0, \qquad \Gamma = \frac{\log|\mathcal{R}(E)|}{2\partial_ES_{01}(E)}.
\end{equation}
where the reflection coefficient is $|\mathcal{R}|=X^{-1}$ with $X$ defined in \eqref{res_open}. 
Note that by assuming a large barrier with $r_1\ll r_2$, one finds that the integral over the barrier scales as $S_{12}\sim E^{-\frac{1}{2}}$ to leading order in $r_1/r_2$, which leads to the Geiger-Nutall rule $\log|\Gamma| \sim E^{-\frac{1}{2}} + ...$.

The likeness of this resonance to the unstable vortex mode become apparent, firstly, by comparing Fig.~\ref{fig:gamow} with the vortex potential on the right panel of Fig.~\ref{fig:comp}.
In both cases, the resonant mode is trapped in a cavity centred on the origin, then tunnels through a barrier out to infinity rendering the eigenvalue complex.
This analogous confinement suggests we can think of the vortex instability as occupying the ``nucleus'' of the vortex.
The similarities are further apparent by comparing the resonance conditions for the $\alpha$-particle in \eqref{alpha_res} to those for the vortex instability in \eqref{res_open2}.
The key difference is that in the case of the $\alpha$-particle there is no amplification mechanism, i.e. $|\mathcal{R}|<1$.
The consequence of this is that $\partial_ES_{01}>0$ and therefore $\Gamma<0$, meaning the $\alpha$-resonance is stable.

\section{Conclusion}

In this work, we have studied the instability of quantised vortices in BECs from the perspective of the WKB approximation.
As well as yielding good agreement with more accurate numerical computations of the instabilities, this method allowed us to explicitly demonstrate that the unstable modes originate inside the ergoregion.
Our analysis also revealed that the instabilities of the open system are recovered when taking the infinite limit of the closed system.
We then studied an analogue black hole by taking our BEC vortex in the non-dispersive approximation and adding a draining component to the flow.
There, we showed that the possibility of superradiant scattering is a result of the same tunnelling of positive norm modes to the negative norm branch of the dispersion relation, which was the mechanism at the heart of the instability in the non-draining case.
Based on these considerations, we also presented a stability condition for a draining vortex in an infinite (open) condensate when the drain is not a perfect absorber.
This condition will be modified in trapped condensates where the amplified modes are reflected back into the vortex, making mode growth possible outside the ergoregion.

In the last part of this work, we considered the BH bomb instability of a (gravitational) Kerr BH and compared this to this vortex instability.
Our method demonstrated that both processes result from the perpetual amplification of superradiant modes, but the location of the cavity occupied by these modes differs.
In the case of a vortex, the cavity is inside the ergoregion whereas for the black hole, it is outside.
This leads to a natural explanation of the differing long term dynamics in the two cases: the ergoregion instability grows inside the vortex and ultimately causes it to fragment, whereas the BH bomb grows outside the ergoregion where it eventually saps all the angular momentum away from the BH.
Inspired by the hydrogenic approximation for the spectrum of BH bomb modes, we also argued that the vortex instability has a similar origin to the nuclear resonances involved in $\alpha$-decay.

By identifying the confinement mechanisms, our results provide deeper insight into the origin of instabilities in rotating systems.
The method we presented could also be used to better understand the properties of draining superfluid vortices, which have been argued to be stable \cite{zezyulin2014stationary,alperin2021multiply}.
Our analysis also has implications for the evolution of vortex states in symmetric trapping geometries.
In particular, in our companion paper \cite{patrick2021origin} we study the fragmentation of a doubly quantised vortex that ensues from the ergoregion instability, finding that nonlinear modulations of the vortex separation can occur due to a back-and-forth energy exchange between sound and vortices.
We also recently studied the propagation of sound around an $\ell=29$ vortex after it decays into a cluster of $\ell=1$ vortices \cite{geelmuyden2021sound}. An analysis of the instability spectrum of the initial vortex, of the kind we detailed here, could be used to gain insight into the resulting vortex configurations.

As a final note, we would like to point out that interesting dynamics resulting from ergoregion instabilities are not unique to vortices in BECs.
In fact, it is known that classical free surface vortices exhibit rotating-polygon instabilities in containers with the correct dimensions.
It has been argued in \cite{mougel2017instabilities} that this instability is also connected to superradiance, or over-reflection as it is called in fluid dynamics.
There, the shape of the polygon is related to the $m$-mode which is unstable and the system size plays the same role in it's ability to stabilise the system as in the case of the closed system studied here.
The similarity between our stability map and those in \cite{mougel2017instabilities} is a strong indication that the same mechanism is at play in both systems and the different dynamics (polygon vs. vortex splitting) can be understood by the different behaviour of the free surface/density profiles.
It would be interesting to apply our analysis to the polygon-instability to understand how the dispersive nature of surface water-waves can influence the ensuing dynamics.

\bibliography{main.bbl}
\bibliographystyle{apsrev4-2}

\appendix
\section{Density of an $\ell$-charged vortex} \label{app:dens_vor}

To solve \eqref{Yeq} for $Y(r)$, we use a shooting method. 
First we rewrite the second order differential equation for $Z$ as two couple first order equations,
\begin{equation} \label{dens_eqs}
\begin{split}
Y' = & \ Z, \\
Z' = & \ - \frac{1}{r}Z - \left(2-\frac{\ell^2}{r^2}\right)Y + 2Y^3,
\end{split}
\end{equation}
with $'=\partial_r$.
We evolve these equations from the starting point $r_1$ using a standard differential equation solver (we used Matlab's inbuilt function \textit{ode45}).
Since setting $r_1=0$ causes a divergence in the equations, we specify our initial conditions a small distance from the origin $r_1=\varepsilon$.
Using the asymptotics in \eqref{Z_asymps}, these are chosen to be $Y(\varepsilon)=Y_0 J_\ell(\varepsilon)$ and $Y'(\varepsilon)=Y_0 J'_\ell(\varepsilon)$.

The value of $Y_0$ is iteratively improved until the solution converges to the desired value at $r_{N_r}$.
Since we are ultimately interested in $Z(r)$ profiles in infinite systems, one would ideally set $r_{N_r}$ to some large value and find the value of $Y_0$ such that $Z_{N}=0$.
However, the shooting problem for \eqref{Yeq} is inherently unstable and if $r_{N}$ is chosen too large then the value of $Y_0$ needs to be specified to an unworkable precision.
Instead, we look for the solution satisfying a Dirichlet BC, $Y_{N}=0$, which corresponds to the wavefunction of a condensate inside a trapping potential of the form $U(r<r_{N})=0$ and $U(r= r_{N})=\infty$.
Once obtained, we then use \eqref{Yeq} to find the leading terms in the Taylor expansion of $Y$ in the limit $r\to\infty$,
\begin{equation} \label{taylor}
\begin{split}
Y(r) \overset{\infty}{\sim} & \ 1 - \frac{\ell^2}{4r^2} - \left(1+\frac{\ell^2}{8}\right)\frac{\ell^2}{4r^4} \\
& \qquad - \left(8+2\ell^2+\frac{\ell^4}{16}\right)\frac{\ell^2}{8r^6} + \mathcal{O}\left(\frac{1}{r^8}\right).
\end{split}
\end{equation}
This is stitched onto the numerical solution using a ramp function over the range $r\in[a_1,b_1]$. 
In order to guarantee the correct limiting behaviour at the origin, we also stitch our numerical solution onto the asymptotic solution $B J_\ell(r)$ at the origin, using a ramp function over the range $r\in[a_2,b_2]$, with the parameter $B$ chosen to best fit the numerical solution in the region $0<r<b_2$.

In our solver, we set the initial value of $Y_0$ to 1.
The effects of different initial points were checked using the values $\varepsilon=10^{-4},10^{-3},10^{-2}$ and no noticeable difference was found.
We set the values $r_N=20$, $a_1=7$ and $b_1=15$ to minimise the time taken to find a convergent solution in the numerics whilst still yielding a good match in the stitching region.
Near the origin, the stitching was performed using $a_2=0.25$ and $b_2=0.5$.

\section{WKB amplitude} \label{app:amp}

In this appendix, we derive the expression for the WKB amplitude in \eqref{WKB2amp}. 
The WKB modes and their $y$ derivatives up to fourth order are,
\begin{equation} \label{WKBmodes}
\begin{split}
f & \ = \Afr e^{i\int q dy}, \\
f' & \ = \left(iq\Afr+\Afr'\right)e^{i\int q dy}, \\
f'' & \ = \left(-q^2\Afr+iq'\Afr+2iq\Afr'\right)e^{i\int q dy} + ... \\
f''' & \ = \left(-iq^3\Afr-3qq'\Afr-3q^2\Afr'\right)e^{i\int q dy} + ... \\
f'''' & \ = \left(q^4\Afr-6iq^2q'\Afr-4iq^3\Afr'\right)e^{i\int q dy} + ...
\end{split}
\end{equation}
where only terms involving up to one derivative on either $q$ or $\Afr$ have been retained, since these are the terms which contribute up to next-to-leading order (NTLO) in the WKB expansion.
Using just the first term from each of \eqref{WKBmodes} gives the leading order (LO) equation in \eqref{disp1}.
The NTLO terms give an equation for the evolution of the amplitude.
To find this equation, we can combine the two equations in \eqref{new_eqs} into a single equation for $f$.
The NTLO contributions are,
\begin{equation}
\begin{split}
& \zeta f''''+2\zeta'f'''-\left(2\m^2+4\rho r^2\right)\zeta f'' \\
& \qquad \qquad \qquad \qquad -2(\zeta\m^2)'f' + ... = 0,
\end{split}
\end{equation}
where $\zeta=1/\Omega r^2$ has been defined for conciseness.
Using the expressions in \eqref{WKBmodes} and extracting the NTLO terms, one finds,
\begin{equation}
\begin{split}
0 = & \ 6\zeta q^2 q'\Afr  + 4\zeta q^3 \Afr' + 2\zeta' q^3 \Afr + 2(\zeta\m^2)'q\Afr \\
& \qquad \qquad \quad + \left(2\m^2+4\rho r^2\right)\zeta\left(q'\Afr+2q\Afr'\right), \\
= & \ \Afr^{-1}\big[\partial_y(2\zeta q^3 \Afr^2) + \partial_y(2\zeta \m^2 q\Afr^2) \\
& \qquad \qquad \quad + 4\rho r^2 \zeta \partial_y(q\Afr^2)\big],
\end{split}
\end{equation}
where in going to from the first to the second line, we have multiplied everything by $\Afr$ and collected terms under the $y$ derivatives. The LO equation in \eqref{disp1} can be written in the form,
\begin{equation}
\zeta^2\kappa^4 + 4\rho r^2\zeta^2\kappa^2 = 4, \qquad \kappa^2 = q^2+\m^2.
\end{equation}
Using this notation and multiplying the equation above by $\kappa^2\zeta \Afr$ we find,
\begin{equation}
\begin{split}
0 = & \ 2\zeta\kappa^2\partial_y(\zeta\kappa^2q\Afr^2) + 4\rho r^2\zeta^2\kappa^2\partial_y(q\Afr^2), \\
= & \ 2\zeta\kappa^2\partial_y(\zeta\kappa^2q\Afr^2)-\zeta^2\kappa^4\partial_y(q\Afr^2)+4\partial_y(q\Afr^2), \\
= & \ \partial_y(\zeta^2\kappa^4 q\Afr^2) + 4\partial_y(q\Afr^2),
\end{split}
\end{equation}
Dividing by 4 and using $q = p r$ and $\kappa = k r$, the right hand side becomes,
\begin{equation}
\begin{split}
\partial_y\left[\left(1+\tfrac{1}{4}\zeta^2\kappa^4\right)q\Afr^2\right] = & \ \partial_y\left[\left(1+\frac{k^2}{4F}\right)rp\Afr^2\right], \\
= & \ r\partial_r\left[\left(\frac{\rho+k^2/2}{\rho+k^2/4}\right)rp\Afr^2\right],
\end{split}
\end{equation}
from which we obtain the result for the amplitude \eqref{WKB2amp} in the main text.

\section{Simulations of the BdG equation} \label{app:BdG_sims}

We evolve the BdG equation \eqref{BdGeq} using a Method of Lines approach.
First, we define a discrete $(r,t)$ grid with spacing $\Delta r=0.4$ and $\Delta t=\Delta r^2/2$. 
We then write $\partial_t|U\rangle = -i\widehat{L}|U\rangle$, where the derivative terms in $\widehat{L}$ (i.e. $\partial_r$ and $\partial_r^2$) are expressed as 5-point (centred) finite difference stencils which encode the correct boundary conditions at $r=0$ (as discussed in Sec.~\ref{sec:asymp}).
We evolve in time using a trapezium method, i.e.
\begin{equation}
|U(t+\Delta t)\rangle = \left(\frac{1-i\widehat{L}\Delta t/2}{1+i\widehat{L}\Delta t/2}\right)|U(t)\rangle,
\end{equation}
where the term on the denominator is understood as a multiplication by the inverse matrix.
For our initial condition, we take $u_+(t=0)=e^{-(r-r_p)^2/2\sigma}$, with $r_p=15$ and $\sigma=2$, and $u_-(t=0)=0$.
The outer boundary (where we impose Dirichlet BCs on both $u_+$ and $u_-$) is located at $r_D=400$, which is sufficiently far away that no reflections re-enter our window of analysis.
To extract the frequency content, we consider the signal $S(t)=u_+(r_c,t)$, where $r_c$ is the point inside the vortex core where the wave amplitude is largest.
We compute the Fourier transform $\tilde{S}(\omega)=\mathscr{F}[H(t)S(t)]$, where $H(t)$ is the Hamming window function which minimises the secondary lobes in $\tilde{S}(\omega)$.
The locations of the peaks of $\tilde{S}(\omega)$, which are obtained via parabolic interpolation, give the real part of the unstable frequency.
To find the imaginary part, we fit a straight line to $\log|S(t)|$ in the region where the instability dominates the signal.
It is not always possible to extract the imaginary part, for example, when two instabilities inference with each other or an instability interferes with a slowly decaying mode.

\section{Klein-Gordon norm} \label{app:norm}

The Klein-Gordon equation in \eqref{KGeqn} can be derived by minimising the following action,
\begin{equation}
\mathcal{S}_\mathrm{KG} = \int d^3\mathbf{x} \sqrt{-g}\left(-\partial^\nu\phi\,\partial_\nu\phi^* - \mu^2\phi\,\phi^*\right).
\end{equation}
Since this invariant under phase rotations of $\phi$, it admits a conserved norm,
\begin{equation} \label{norm_kerr}
j^\nu = \sqrt{-g}g^{\nu\sigma}i\left(\phi\,\partial_\sigma\phi^*-\phi^*\partial_\sigma\phi\right),
\end{equation}
satisfying the conservation equation $\partial_\nu j^\nu=0$.
Using the decomposition of $\phi$ in \eqref{kerr_decomp} the main text with the angular functions normalised as,
\begin{equation}
2\pi\int^\pi_0\sin\theta\,S^2_{\omega m}(\theta)\,d\theta = 1,
\end{equation}
the $j^r$ component is fixed with $r$, leading to,
\begin{equation}
\partial_r\left( \Delta_r[R_{\omega m}\partial_r R^*_{\omega m} - R^*_{\omega m}\partial_r R_{\omega m}] \right) = 0,
\end{equation}
where the quantity in round parentheses is the conserved Wronskian associated with \eqref{Kerr_R_eq}.

The time component of \eqref{norm_kerr} $j^t$ is the norm (density).
In Section~\ref{sec:wkb_method} it was possible to find a concise expression for the norm of the vortex excitations using the WKB approximation, which then allowed us to deduce the sign of the norm in different regions of the flow.
Things are not as simple here, since the $\theta$ dependence in the $S_{\omega m}$ functions contained in $\phi$ complicates the expressions.
Instead, we consider the following argument.

First, we plug the metric functions \eqref{Kerr} and the $\phi$ decomposition \eqref{kerr_decomp} into the expression for $j^t$ which results in,
\begin{equation} \label{jt}
\begin{split}
j^t = & \ \frac{2}{\Delta_r}\Big[\left\{(r^2+a^2)^2-\Delta_ra^2\sin^2\theta\right\}\omega \\
& \qquad\quad - am\left(r^2+a^2-\Delta_r \right)\Big]\sin\theta\,|\phi|^2.
\end{split}
\end{equation}
Now consider that we have a mode with $\omega>\mu$ so that the mode propagates at spatial infinity,
\begin{equation}
R_{\omega m} \to \frac{A^\pm_\infty}{\sqrt{r^2+a^2}}e^{\pm i\sqrt{\omega^2-\mu^2}r}.
\end{equation}
Inserting this into \eqref{jt} an integrating a wave packet localised in the asymptotic region gives the expression,
\begin{equation}
\int d^3\mathbf{x}\, j^t(\omega>\mu) \sim 2\omega |A_\infty^\pm|^2.
\end{equation}
Therefore, positive frequency propagating modes have positive norm at infinity.
However, in the WKB approximation, the norm of a mode is conserved unless we cross a turning point where there is an exchange of norm between modes.
Therefore, the $\omega>\mu$ modes have positive norm everywhere to the right of the potential barrier in the left panel of Fig.~\ref{fig:comp}.
We can then extend this argument to lower frequencies.
Starting just above $\omega>\mu$, we can decrease the frequency into the region where $\omega<\mu$ and provided we don't cross any turning points, these modes will be positive norm also.
Therefore, the Kerr bound states that exist in the cavity outside the BH have positive norm.

Now we perform a similar analysis for modes which cross the horizon.
Near the horizon, \eqref{jt} becomes,
\begin{equation}
j^t \overset{r_h}{\sim} \frac{2(r^2+a^2)^2}{\Delta_r}\tilde{\omega}_h \sin\theta\,|\phi|^2.
\end{equation}
Then we insert the asymptotic expression from \eqref{R_asymp} and integrate $j^t$ for a wave packet localised at the horizon to find,
\begin{equation}
\int d^3\mathbf{x}\, j^t \sim 2\tilde{\omega}_h |A_h|^2.
\end{equation}
Therefore, modes with $\tilde{\omega}_h<0$ (i.e. the superradiant ones) have negative norm at the horizon whereas those with $\tilde{\omega}_h>0$ have positive norm.
Then applying the WKB approximation, the sign of the norm is unchanged everywhere to the left of the barrier in the left panel of Fig.~\ref{fig:comp}.
Thus, we find that the instabilities (which necessarily have $\tilde{\omega}_h<0$) have negative norm when they emerge on the inside of the barrier and dissipate into the horizon.

\section{Separation constant} \label{app:sep}

In this appendix, the computation of the separation constant $j_{\omega m}$ is described.
First, it is estimated within the WKB approximation, in both flat space and the Kerr spacetime, which provides an intuitive explanation of the value of $j_{\omega m}$ as an ``angular energy level''.
Next, an approximation for $j_{\omega m}$ is provided for $\omega\sim\mu$.
Finally, a numerical procedure is outlined to determine the precise value.

\subsection{WKB method}

The method will involve writing \eqref{S_eq} in the form of a Schr\"odinger equation,
\begin{equation} \label{ang3}
\frac{1}{\cos\chi}\partial_\chi\left(\cos\chi\partial_\chi S_{\omega m}\right) = \left[Q(\chi)-j_{\omega m}\right]S_{\omega m},
\end{equation}
which allows us to interpret $j_{\omega m}$ as an ``energy level'' and $Q(\chi)$ as a potential, where,
\begin{equation}
Q(\chi) = \mu^2 a^2\sin^2\chi + \frac{(\omega a\cos^2\chi-m)^2}{\cos^2\chi}.
\end{equation}
We have also redefined the angular coordinate $\chi=\theta-\pi/2$, which has the range $\chi\in[-\pi/2,\pi/2]$, so that $Q$ is symmetric in $\chi$.
The WKB solutions to \eqref{ang3} are,
\begin{equation} \label{S_wkb}
S_{\omega m} \sim  \left|\cos\chi\sqrt{j_{\omega m}-Q}\right|^{-\frac{1}{2}} e^{\pm i\int\sqrt{j_{\omega m}-Q}\,d\chi}.
\end{equation}
For $|m|\geq 1$, we have $Q\to\infty$ on the poles, the solutions there will be evanescent.
Since the limit of \eqref{ang3} on the poles is Bessel's equation, we expect one solution to be regular and one to be divergent.
To show this, first note that approaching the poles the wavevector goes as $\sqrt{j_{\omega m}-Q}\sim i|m|\sec\chi$.
The amplitude becomes a constant and the phase is,
\begin{widetext}
\begin{equation}
\exp\left(\pm i\int d\chi\sqrt{j_{\omega m}-Q}\right)\sim \exp\left(\mp\int d\chi |m|\sec\chi\right) = \exp\left[\mp 2|m|\tanh^{-1}\tan\left(\chi/2\right)\right].
\end{equation}
Then writing,
\begin{equation}
2\tanh^{-1}\tan(\chi/2) = \log\left(1+\tan\frac{\chi}{2}\right)-\log\left(1-\tan\frac{\chi}{2}\right),
\end{equation}
and also using,
\begin{equation}
\tan\frac{\chi}{2}\sim\begin{cases}
1 + (\chi-\pi/2), \qquad \chi\sim\pi/2, \\
-1 + (\chi+\pi/2), \quad \ \chi\sim-\pi/2,
\end{cases}
\end{equation}
we find,
\begin{equation} \label{asymp}
\exp\left[\mp 2|m|\tanh^{-1}\tan\left(\chi/2\right)\right] \sim \begin{cases}
(\pi/4-\chi/2)^{\pm|m|}, \quad \chi\sim\pi/2, \\
(\pi/4+\chi/2)^{\mp|m|}, \quad \chi\sim-\pi/2.
\end{cases}
\end{equation}
\end{widetext}
The $\chi$ dependence agrees exactly with the limiting form of the Bessel functions as expected.
Hence, there is one divergent mode and one regular mode at each pole.
For the solution to be regular, we must discard the divergent modes.

Now we perform a scattering computation, using the methods of Section~\ref{sec:scatter}, to find a condition for the $j_{\omega m}$ such that the solution is regular.
The potential $Q$ (which is symmetric) will either have two or four turning points, which are the locations where $Q(\chi_\tp)=j_{\omega m}$.
For the case where there are two turning points located at $\pm\chi_1$, we write,
\begin{equation} \label{2tp_mat_eq}
\begin{pmatrix}
A^\uparrow_{-1} \\ A^\downarrow_{-1}
\end{pmatrix} = \widetilde{T}\begin{pmatrix}
e^{-iS_{-1,+1}} & 0 \\ 0 & e^{iS_{-1,+1}}
\end{pmatrix} T \begin{pmatrix}
A^\downarrow_{+1} \\ A^\uparrow_{+1},
\end{pmatrix},
\end{equation}
where the $A$ are the amplitudes of the WKB modes in \eqref{S_wkb}, the subscript $\pm 1$ is used to indicates a quantity is evaluated at $\pm\chi_1$, and $\uparrow$ ($\downarrow$) denotes the mode which grows (decays) with increasing $\chi$.
The phase term is given by,
\begin{equation}
S_{ij} = \int^{r_j}_{r_i}\left|j_{\omega m} - Q\right|^\frac{1}{2}d\chi.
\end{equation}
From the discussion above, the regularity requirement at the poles leads to the BCs $A^\downarrow_{-1}=A^\uparrow_{+1}=0$.
Then, \eqref{2tp_mat_eq} can be solved to obtain the condition,
\begin{equation} \label{2tp_cond}
S_{-1,+1} = \pi\left(q+\tfrac{1}{2}\right),
\end{equation}
where $q=0,1...$ indexes the different ``energy levels''. 
In the case of four turning points, which we label $-\chi_1<-\chi_0<+\chi_0<+\chi_1$, the scattering computation is,
\begin{widetext}
\begin{equation} \label{4tp_mat_eq}
\begin{pmatrix}
A^\uparrow_{-1} \\ A^\downarrow_{-1}
\end{pmatrix} = \widetilde{T}\begin{pmatrix}
e^{-iS_{-1,-0}} & 0 \\ 0 & e^{iS_{-1,-0}}
\end{pmatrix} T \begin{pmatrix}
0 & e^{S_{-0,+0}} \\ e^{-S_{-0,+0}} & 0
\end{pmatrix} \widetilde{T} \begin{pmatrix}
e^{-iS_{+0,+1}} & 0 \\ 0 & e^{iS_{+0,+1}}
\end{pmatrix} T \begin{pmatrix}
A^\downarrow_{+1} \\ A^\uparrow_{+1}
\end{pmatrix}.
\end{equation}
\end{widetext}
Since $Q(\chi)$ is symmetric, we have $S_{-1,-0}=S_{+0,+1}$.
Imposing $A^\downarrow_{-1}=A^\uparrow_{+1}=0$ and solving \eqref{4tp_mat_eq} gives the condition,
\begin{equation} \label{4tp_cond}
4\cot^2 S_{+0,+1} = e^{-2S_{-0,+0}}
\end{equation}
Note that one also obtains the conditions in \eqref{2tp_cond} and \eqref{4tp_cond} for the case where $m=0$ using an argument similar to that presented in Section~\ref{sec:meql}, that is, by matching the asymptotic behaviour of the exact solution at the poles onto the WKB modes to obtain a relation between the amplitudes.
In that case, the amplitudes of propagating modes at the poles obey the same relations as those at the turning points in the $|m|\geq 1$ case.
It is interesting to note then that the correct BC at the poles for $m=0$ can be formally obtained by taking the $m\to 0$ limit of the $|m|\geq 1$ result, which has the effect of sending the turning points to the poles.

\begin{figure*} 
\centering
\includegraphics[width=.8\linewidth]{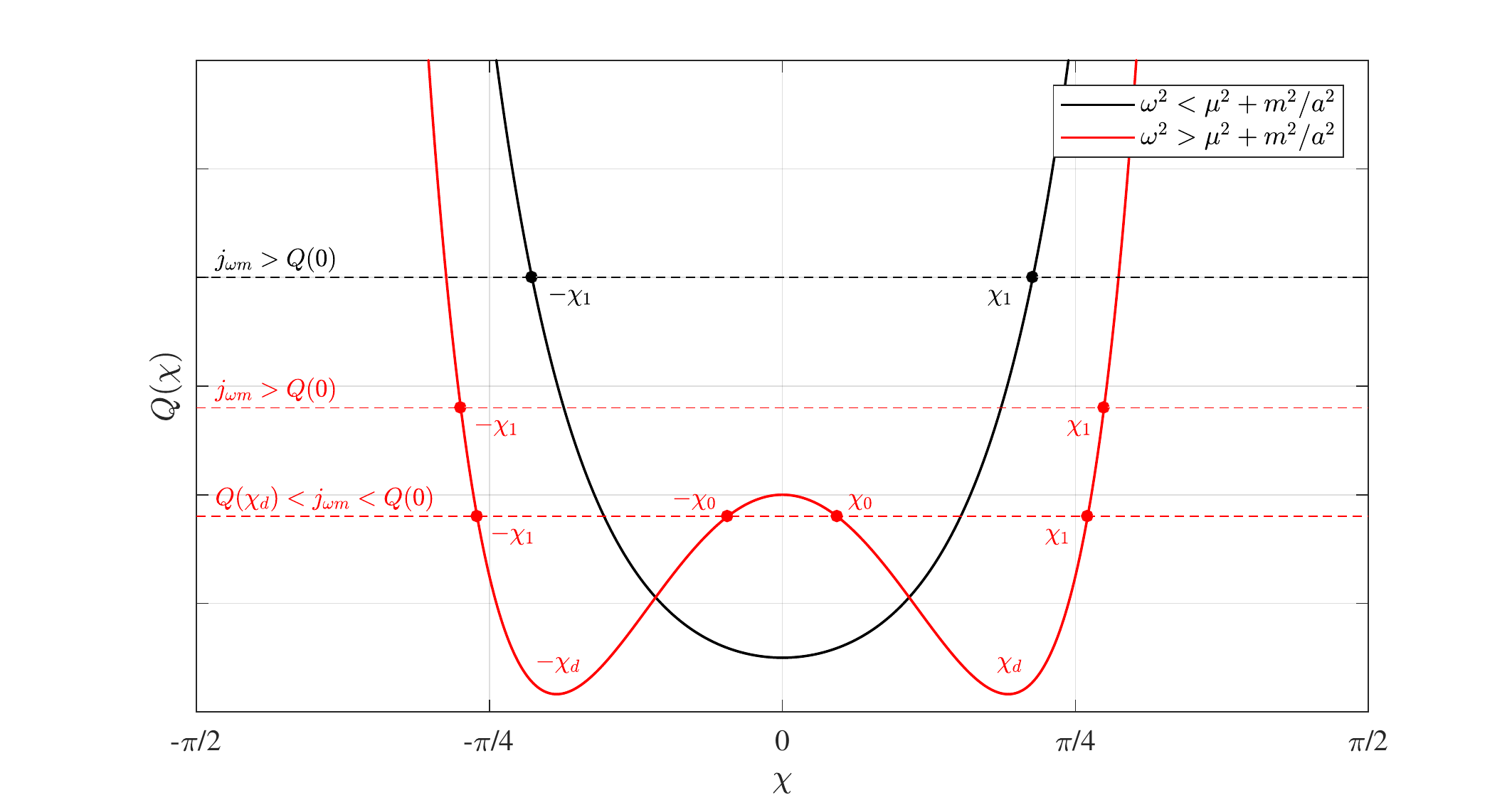}
\caption{Examples of the function $Q(\chi)$.
Interpreting the eigenvalue $j_{\omega m}$ as an ``energy level'' and $Q$ as the potential, the angular equation can be interpreted as a Schr\"odinger equation for a particle with wavefunction $S$.
Regions where $j_{\omega m}>Q$ are classically allowed (i.e. waves propagate) whereas those with $j_{\omega m}<Q$ are classically forbidden (i.e. waves are evanescent).
Three possibilities are indicated for the location of the turning points and the extrema of $Q$.
} \label{fig:Q_pot}
\end{figure*}

\begin{figure*} 
\centering
\includegraphics[width=\linewidth]{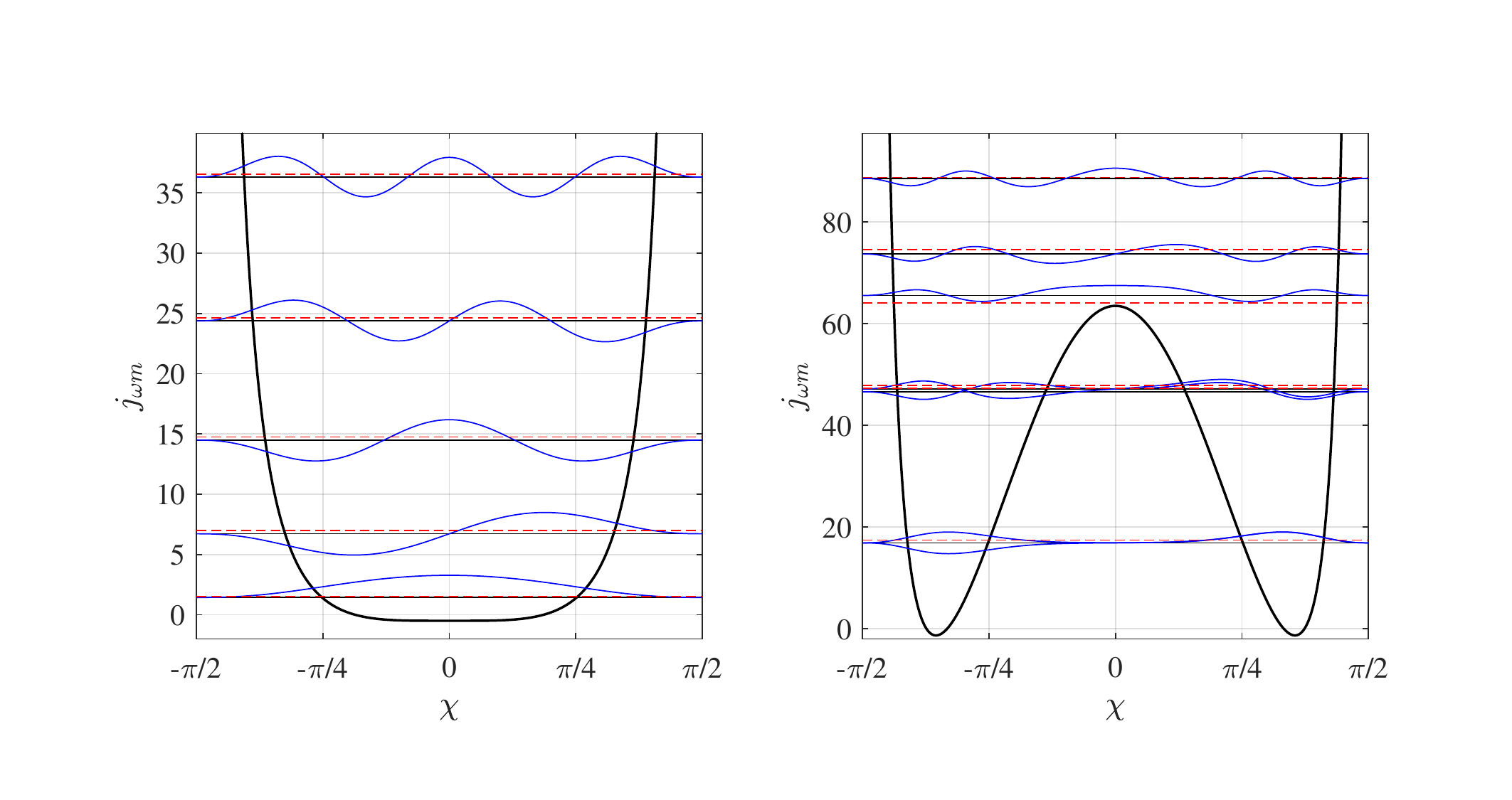}
\caption{Comparison of the $j_{\omega m}$ obtained from a numerical calculation (horizontal black line) and a WKB approximation (horizonal red dashed line) for the parameters $a=0.5$, $m=2$, $\mu=1$.
The frequency is $\omega=4$ on the left and $\omega=20$ on the right.
The two results are in good agreement except near the peak of the central barrier where the WKB method is expected to fail.
Also displayed is the potential appearing in the numerical calculation, i.e. $Q-(\sec^2\chi+1)/4$ (black curve), and the wavefunctions $v$ are superimposed over the eigenvalues (see Appendix~\ref{app:angEV_num} for details).
} \label{fig:J_levs}
\end{figure*}

\subsubsection{Application to flat space}

In the case of flat space, we have $a=0$ and the potential simplifies to $Q=m^2/\cos^2\chi$.
There are two turning points $\pm\chi_1 = \pm\cos^{-1}\lambda$ with $\lambda=|m|/\sqrt{j_{\omega m}}$.
We can then write the integral in \eqref{2tp_cond} as $S_{-1,+1} = \sqrt{j_{\omega m}}\,\mathcal{I}(\lambda)$, where,
\begin{equation}
\begin{split}
\mathcal{I}(\lambda) = & \ \int^{\cos^{-1}\lambda}_{-\cos^{-1}\lambda}d\chi\sqrt{1-\frac{\lambda^2}{\cos^2\chi}}, \\
= & \ \Bigg[\sin^{-1}\left(\frac{\sin\chi}{\sqrt{1-\lambda^2}}\right) \\
& \quad -\lambda\tan^{-1}\left(\frac{\lambda\sin\chi}{\sqrt{\cos^2\chi-\lambda^2}}\right)\Bigg]^{\cos^{-1}\lambda}_{-\cos^{-1}\lambda} \\
= & \ \pi(1-\lambda).
\end{split}
\end{equation}
Using this result, we can solve \eqref{2tp_cond} for the separation constant,
\begin{equation}
j_{\omega m} = \left(l+\tfrac{1}{2}\right)^2,
\end{equation}
with $l=|m|+q\geq|m|$. 
This differs from the exact value of $j_{\omega m}$ for the spherical harmonics, i.e. $l(l+1)$, only by 1/4.
This difference quickly becomes small for large $l$, and the approximation is still satisfactory even for $l=1$.

\subsubsection{Application to Kerr}

In the Kerr spacetime, we have $a\neq 0$.
In this case, function $Q(\chi)$ has extrema at the following locations,
\begin{equation}
\chi = 0,\pm\chi_d, \qquad \chi_d = \cos^{-1}\sqrt{\frac{|m|}{a\sqrt{\omega^2-\mu^2}}},
\end{equation}
which distinguishes two possibilities for the shape of $Q$:
\begin{itemize}
\item If $\omega^2<\mu^2+m^2/a^2$ then $Q$ has a single extremum at $\chi=0$ which is a minimum.
Modes with $j_{\omega m}>Q(0)$ have two turning points at located at $\pm\chi_1$. 
\item If $\omega^2>\mu^2+m^2/a^2$ then $Q$ has a maximum at $\chi=0$ and minima at $\chi=\pm\chi_d$.
Modes with $j_{\omega m}>Q(0)$ have two turning points $\pm\chi_1$ like the previous case.
However, for $Q(\chi_\pm)<j_{\omega m}<Q(0)$ there will be four turning points $\pm\chi_0,\pm\chi_1$. 
\end{itemize}
The various scenarios are illustrated in Fig.~\ref{fig:Q_pot}.

In the case of two turning points, we need to solve \eqref{2tp_cond} whereas for four turning points we solve \eqref{4tp_cond}.
We integrate the integrals appearing in these equations numerical and plot the resulting eigenvalues as red dashed lines in Fig.~\ref{fig:J_levs}.
The WKB results are in good agreement with the exact values (evaluated in Appendix~\ref{app:angEV_num}), except near the peak of the central barrier where the $\pm\chi_0$ become too close to each other and the WKB approximation fails in between. 

We can also use the WKB results to see why modes deep in the double well of $Q$ begin to overlap.
Starting from the condition in \eqref{4tp_cond}, we may write,
\begin{equation} \label{barrier_Kerr_ang}
S_{+0,+1} = \pi\left(q+\tfrac{1}{2}\right) \pm \cot^{-1}e^{-S_{-0,+0}},
\end{equation}
where, here, $q=0,1...$ only counts half of the available energy levels since there are now two conditions.
For a mode deep in the well, the barrier is large and the exponential in this expression will be very small.
We can then approximate $\cot^{-1}e^{-S_{-0,+0}}\approx e^{-S_{-0,+0}}$.
In this case, there is only a small difference in the $j_{\omega m}$ which solve the two conditions.
Hence, there will be almost overlapping modes deep in the well which are out of phase on one side.

\subsection{Approximation for $\omega\sim\mu$}

For $\omega\sim\mu$, the term proportional to $\mu^2-\omega^2$ in \eqref{S_eq} can be neglected and the angular equation becomes,
\begin{equation}
\sin\theta\partial_\theta\left(\sin\theta\partial_\theta S\right) = \left( m^2 - A_{\omega m}\sin^2\theta\right) S,
\end{equation}
with $A_{\omega m} = j_{\omega m} - a\omega(a\omega-2m)$.
The eigenvalue of this equation is exactly $A_{\omega m}=l(l+1)$.
Thus, we have,
\begin{equation} \label{j_approx}
j_{\omega m}(\omega\sim\mu) = l(l+1) + a\omega(a\omega-2m).
\end{equation}

\begin{figure} 
\centering
\includegraphics[width=\linewidth]{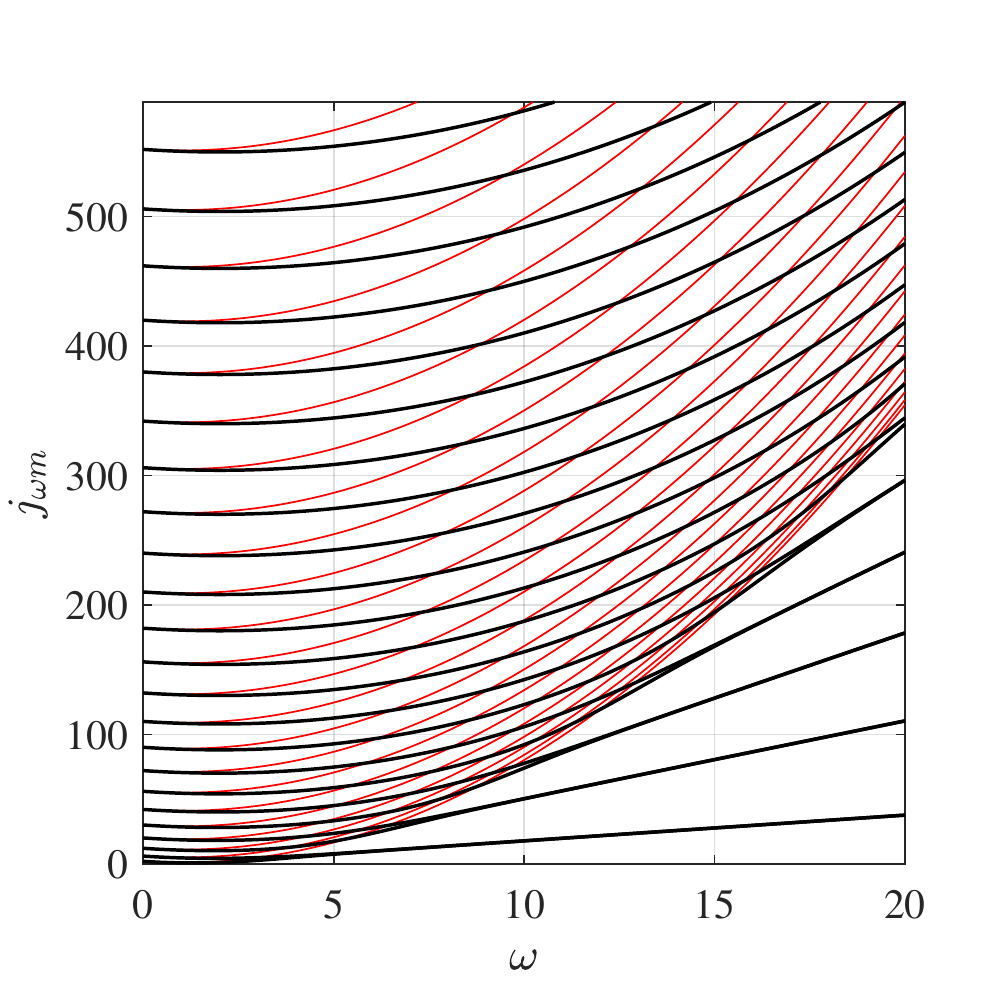}
\caption{The dependence of the separation constant $j_{\omega m}$ on $\omega$ for the parameters $m=1$, $a=0.99$ and $\mu=0.42$.
The black curves represent the numerically computed values (with the lowest $l$ at the bottom) and the red curves are the approximation in \eqref{j_approx}, which are in good agreement with the numerical values $\omega\sim\mu$.
The change in behaviour in the bottom right hand corner corresponds to states below the peak of the central barrier in the $Q$ potential. 
As $\omega$ increases, the states move deeper into the two wells and neighbouring eigenvalues merge as predicted by \eqref{barrier_Kerr_ang}.
} \label{fig:sep}
\end{figure}

\subsection{Numerical method} \label{app:angEV_num}
For numerical evaluation it is easier to work with \eqref{S_eq} after defining a new field $v=S\sqrt{\sin\theta}$, which puts the angular equation into the form,
\begin{equation}
\widehat{\mathcal{D}}v = j_{\omega m} v,
\end{equation}
where the differential operator is,
\begin{equation}
\widehat{\mathcal{D}} = Q(\chi) -(1+\sec^2\chi)/4 - \partial_\chi^2,
\end{equation}
and the $j_{\omega m}$ and $v$ are it's eigenvalues and eigenfunctions respectively.
Looking at the asymptotic behaviour of $S$ in \eqref{asymp}, we see that (for $|m|\geq1$) we need to impose Dirichlet BCs at the poles, i.e. $v(\chi=\pm\pi/2)=0$.
We can then write $\widehat{\mathcal{D}}$ as a finite difference matrix which encodes the correct BCs and search for it's eigenvalues, which are the $j_{\omega m}$.
The first few eigenvalues and their associated eigenfunctions are illustrated in Fig.~\ref{fig:J_levs}.
The dependence of $j_{\omega m}$ on $\omega$ is illustrated in Fig.~\ref{fig:sep}.

\section{Hydrogenic spectrum}

In this appendix, a derivation of the hydrogenic spectrum from the resonance condition in \eqref{BohrSomm} is provided.
Firstly, the effective potential $-K^2$ is a quartic polynomial which can be written in the form,
\begin{equation}
\begin{split}
-K^2 = & \ \bm{a}r^4 + \bm{b}r^3 + \bm{c}r^2 + \bm{d}r + \bm{e}, \\
\bm{a} = & \ \mu^2-\omega^2, \quad \bm{b} = -2\mu^2, \quad \bm{d} = -2j_{\omega m}, \\
\bm{c} = & \ 2\omega ma-2\omega^2a^2+j_{\omega m}+\mu^2a^2, \\
\bm{e} = & \ 2\omega ma^3-\omega^2a^4-m^2a^2+j_{\omega m}a^2,
\end{split}
\end{equation}
where bold letters are the polynomial coefficients.
When $\omega\to\mu$, $r_3$ diverges as in \eqref{r3_div} and the integral for $I$ can be approximated in the following manner.
For large $r$, $-K^2$ is well approximated by keeping only the quartic, cubic and quadratic terms, and we also have $\Delta_r\sim r^2$. 
Thus, the integral can be written,
\begin{widetext}
\begin{equation} \label{int1}
\begin{split}
I\approx & \ \int^{r_3}_{r_2}\frac{dr}{r^2} \widetilde{K}, \qquad \quad -\widetilde{K}^2 = \bm{a}r^4+\bm{b}r^3+\bm{c}r^2, \\
= & \ \Bigg[\frac{\widetilde{K}}{r} + \bm{c}^\frac{1}{2}\tan^{-1}\left(\frac{2\bm{c}r+\bm{b}r^2}{2\bm{c}^\frac{1}{2}\widetilde{K}}\right) - \frac{\bm{b}}{2\bm{a}^\frac{1}{2}}\tan^{-1}\left(\frac{2\bm{a}r^2+\bm{b}r}{2\bm{a}^\frac{1}{2}\widetilde{K}}\right)\Bigg]^{r_3}_{r_2} = -\pi\left(\frac{\bm{b}}{2\bm{a}^\frac{1}{2}}+\bm{c}^\frac{1}{2}\right)
\end{split}
\end{equation}
\end{widetext}
where $r_{2,3}$ are approximated as the zeros of $-\widetilde{K}^2=0$, which for small $\bm{a}$ are approximately given by,
\begin{equation}
r_2 \approx -\bm{c}/\bm{b}, \qquad r_3 \approx -\bm{b}/\bm{a}.
\end{equation}
Using the $\omega\sim\mu$ approximation for $j_{\omega m}$ in \eqref{j_approx}, we see that $\bm{c}$ approaches a value of $l(l+1)$.
Inserting the expressions for $\bm{a}$, $\bm{b}$ and $\bm{c}$, this gives,
\begin{equation} \label{hydro2}
I(\omega) \approx \frac{\pi\mu^2}{\sqrt{\mu^2-\omega^2}}-\pi\left(l+\tfrac{1}{2}\right),
\end{equation}
where we have used the fact that WKB is approximation formally valid in the limit of large $l$ to write $\sqrt{l(l+1)}=l+\tfrac{1}{2}+\mathcal{O}(l^{-1})$.
This is the result in the main text \eqref{hydro1}.

\end{document}